\documentclass[journal]{IEEEtran}
\usepackage{amsmath,amsfonts}
\usepackage{algorithmic}
\usepackage{algorithm}
\usepackage{array}
\usepackage[caption=false]{subfig}
\usepackage{url}
\usepackage{graphicx}
\usepackage{cite}
\usepackage{colortbl,booktabs} 
\usepackage[table,xcdraw]{xcolor}
\usepackage[normalem]{ulem}
\usepackage{multirow}
\usepackage{balance}
\usepackage{amsmath}
\usepackage[utf8]{inputenc}
\definecolor{teal}{RGB}{62, 153, 159}          % Teal Blue
\definecolor{lightred}{RGB}{255, 153, 153}    % Light Pink
\definecolor{grassgreen}{RGB}{157, 217, 124}     % grass green
\definecolor{darkgrayblue}{RGB}{150, 161, 171}  % Steel Gray
\definecolor{skyblue}{RGB}{135, 206, 235}     % Sky Blue
\definecolor{darkpurple}{RGB}{183, 137, 229}     % Lavender Purple
\usepackage{tikz}
\usepackage[colorlinks,linkcolor=blue,anchorcolor=blue,citecolor=blue,urlcolor=blue]{hyperref}

\makeatletter

\newcommand{\Rmnum}[1]{\expandafter\@slowromancap\romannumeral #1@}
\makeatother

\definecolor{lime}{HTML}{A6CE39}
\DeclareRobustCommand{\orcidicon}{
\begin{tikzpicture}
\draw[lime, fill=lime] (0,0)
circle[radius=0.16]
node[white]{{\fontfamily{qag}\selectfont \tiny \.{I}D}};
\end{tikzpicture}
\hspace{-2mm}
}
\foreach \x in {A, ..., Z}{%
\expandafter\xdef\csname orcid\x\endcsname{\noexpand\href{https://orcid.org/\csname orcidauthor\x\endcsname}{\noexpand\orcidicon}}
}

\begin{document}
\title{BreathNet: Generalizable Audio Deepfake Detection via Breath-Cue-Guided Feature Refinement}

\author{Zhe Ye\hspace{-1.5mm}\orcidA{}\hspace{-1.5mm}, Xiangui Kang\hspace{-1.5mm}\orcidB{}\hspace{-1.5mm},~\IEEEmembership{Senior Member,~IEEE,} Jiayi He, Chengxin Chen, Wei Zhu, Kai Wu, Yin Yang, \\ and Jiwu Huang\hspace{-1.5mm}\orcidC{}\hspace{-1.5mm},~\IEEEmembership{Fellow,~IEEE}

\thanks{Zhe Ye and Xiangui Kang are with the Guangdong Key Laboratory of Information Security, School of Computer Science and Engineering, Sun Yat-sen University, Guangzhou 510006, China  (e-mail:yezh57@mail2.sysu.edu.cn; isskxg@mail.sysu.edu.cn). 

Jiayi He is with the State Key Laboratory of Multi-modal Artificial Intelligence Systems, Institute of Automation, Chinese Academy of Sciences, Beijing 100190, China (e-mail: jiayi.he@ia.ac.cn)

Chengxin Chen, Wei Zhu, Kai Wu, and Yin Yang are with the China Mobile Internet Corporation, China Mobile Communications Corporation, Guangzhou 510630, China (e-mail: \{chenchengxin, zhuwei6, wukai6, yangyin\}@cmic.chinamobile.com). 

Jiwu Huang is with the Guangdong Laboratory of Machine Perception and Intelligent Computing, Faculty of Engineering, Shenzhen MSU-BIT University, Shenzhen 518116, China. (e-mail: jwhuang@smbu.edu.cn)

}}

\markboth{Journal of \LaTeX\ Class Files,~Vol.~XX, No.~X, August~XXXX}
{Ye \MakeLowercase{\textit{et al.}}: Bare Demo of IEEEtran.cls for IEEE Journals}


\maketitle

\begin{abstract}
As deepfake audio becomes more realistic and diverse, developing generalizable countermeasure systems has become crucial. Existing detection methods primarily depend on XLS-R front-end features to improve generalization. Nonetheless, their performance remains limited, partly due to insufficient attention to fine-grained information, such as physiological cues or frequency-domain features. In this paper, we propose BreathNet, a novel audio deepfake detection framework that integrates fine-grained breath information to improve generalization. Specifically, we design BreathFiLM, a feature-wise linear modulation mechanism that selectively amplifies temporal representations based on the presence of breathing sounds. BreathFiLM is trained jointly with the XLS-R extractor, in turn encouraging the extractor to learn and encode breath-related cues into the temporal features. Then, we use the frequency front-end to extract spectral features, which are then fused with temporal features to provide complementary information introduced by vocoders or compression artifacts. Additionally, we propose a group of feature losses comprising Positive-only Supervised Contrastive Loss (PSCL), center loss, and contrast loss. These losses jointly enhance the discriminative ability, encouraging the model to separate bona fide and deepfake samples more effectively in the feature space. Extensive experiments on five benchmark datasets demonstrate state-of-the-art (SOTA) performance. Using the ASVspoof 2019 LA training set, our method attains 1.99\% average EER across four related eval benchmarks, with particularly strong performance on the In-the-Wild dataset, where it achieves 4.70\% EER. Moreover, under the ASVspoof5 evaluation protocol, our method achieves an EER of 4.94\% on this latest benchmark.

\end{abstract}


\section{Introduction}
\IEEEPARstart{R}{ecent} advancements in text-to-speech (TTS) \cite{mehta2024matcha, guan2024mm} and voice conversion (VC) \cite{yao2024promptvc, li2024sef} have enabled the generation of deepfake speech that closely resembles natural human speech in terms of timbre, prosody, speaking rate, and emotional nuance. These AI-generated voices are often indistinguishable from bona fide speech, making them highly deceptive and introducing significant security risks. Malicious attackers can exploit these technologies to synthesize speech that impersonates legitimate users, thereby circumventing speaker verification systems to facilitate fraudulent activities \cite{yuan2025evilharmony, ye23_interspeech, cheng2024alif}. These threats pose serious challenges to the integrity of voice-based authentication systems and broader trust in voice communication. Consequently, the development of high-accuracy and generalization audio deepfake detection techniques has become a critical priority in speech security research \cite{wang24_asvspoof,10.1145/3714458}, as it is essential for protecting biometric systems and ensuring the safe application of voice technologies across high-stakes domains.

\begin{figure}[t]
\centering
\includegraphics[width=0.5\textwidth]{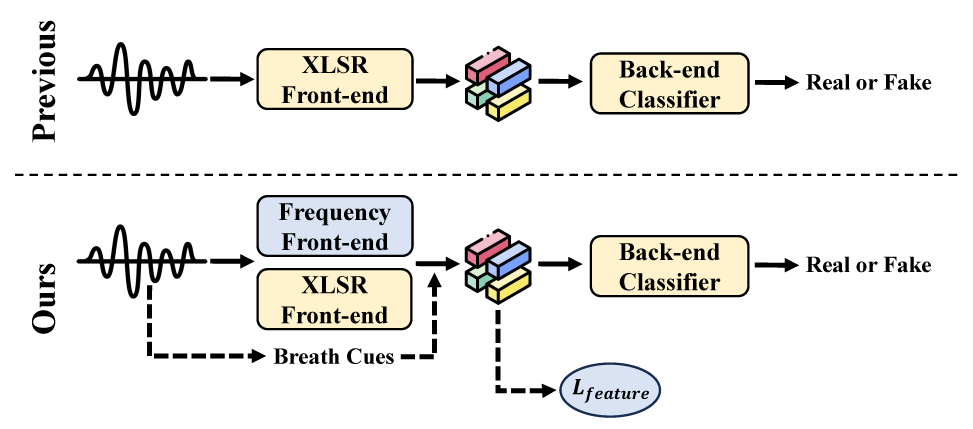}
\caption{Previous methods mainly utilize XLS-R features for detection. Our method integrates spectral features and breath cues, enhancing the discriminative power of the XLS-R features. Besides, our method utilizes feature loss to improve generalization.}
\label{fig-compare}
\end{figure}

Traditional deepfake audio detection methods typically rely on hand-crafted features or deep features. However, such features often exhibit limited generalization ability. Significant performance improvements have been achieved through the use of wav2vec as a feature extractor \cite{hemlata_wav2vec2}. These pre-trained self-supervised models \cite{NEURIPS2020_92d1e1eb, chen2022wavlm}, trained on massive amounts of unlabeled audio, can capture intricate acoustic and linguistic patterns, making them highly effective as feature extractors for various downstream tasks. The proposal of this method marks a significant shift, with large-scale self-supervised features replacing conventional features as the dominant choice.

Despite the successes of using self-supervised features for audio deepfake detection, generalization remains a significant limitation of existing methods. Existing methods rely heavily on self-supervised temporal features while overlooking the importance of frequency-domain features. Artifacts introduced by vocoders and compression, commonly embedded in the frequency domain, can be powerful discriminative information for detecting deepfake audio. Moreover, fine-grained physiological cues, such as breathing sounds, can serve as critical indicators of genuine human speech, yet are also overlooked. Besides, the lack of effective feature refinement hinders the separation between bona fide and deepfake samples in feature space, resulting in limited generalization. 

In this paper, we propose BreathNet, an audio deepfake detection framework that enhances discriminability through breath cues, temporal–spectral feature fusion, and feature refinement, as shown in Fig. \ref{fig-compare}. Specifically, we propose BreathFiLM, which integrates frame-level breath cues into the temporal feature. Serving as an auxiliary training module, it guides the fine-tuned XLS-R to capture breath-related information that is often missing in deepfake speech, enabling the XLS-R to implicitly encode such cues at inference without relying on the breath mask. Next, we fuse the temporal features with spectral features, which enables the model to exploit complementary cues from the frequency domain, improving its ability to detect inconsistencies in deepfake audio. Finally, we propose a group of feature losses where each loss plays a distinct yet complementary role. The PSCL encourages bona fide samples to exhibit sufficient feature similarity. The center loss pulls bona fide samples toward a learnable class center, thereby reducing intra-class variance. PSCL and the center loss jointly ensure intra-class compactness. Besides, the contrast loss pushes deepfake samples away from the bona fide center, guaranteeing inter-class separability and sharpening the decision boundary. Overall, these losses enhance the separability of deepfake and bona fide samples in the feature space, thereby improving generalization.

Extensive experiments on five representative benchmarks, including ASVspoof 2019 LA (19LA) \cite{wang2020asvspoof}, ASVspoof 2021 LA (21LA), ASVspoof 2021 DF (21DF) \cite{liu2023asvspoof}, In-the-Wild (ITW) \cite{muller22_interspeech}, and ASVspoof5 \cite{WANG2026101825}, demonstrate the effectiveness and generalization capability of our method. Ablation studies confirm that each proposed module makes an individual contribution to performance gains, while the integrated system achieves SOTA performance on the challenging ITW dataset. Additionally, we provide t-SNE visualizations of the learned hidden features to reveal the separability of representations and to analyze evaluation strategies for effectiveness. 

The key contributions of our work can be summarized as follows:

\begin{itemize}

\item[$\bullet$] We propose BreathNet, a novel framework for detecting deepfake audio, which explicitly integrates temporal features with breath-related cues and complementary spectral information, enabling the model to capture more discriminative information for deepfake detection.

\item[$\bullet$] We design a group of feature loss where each component plays a distinct role: the PSCL coarsely improves bona fide samples’ similarity, the center loss enhances bona fide compactness, and the contrast loss pushes deepfakes away from the bona fide center, enabling fine-grained separation and improved generalization.

\item[$\bullet$] Extensive experiments on five benchmark datasets validate the effectiveness and generalization of our method. Our method demonstrates SOTA performance, reducing the EER to 4.70\% on ITW and 4.94\% on ASVspoof5.
\end{itemize}

The rest of the paper is organized as follows. We provide a brief review of related works in Section \ref{Related}. Our methodology is described in Section \ref{Methodology}, which is followed by experimental results and analysis in Section \ref{Results}. Finally, conclude this paper in Section \ref{Conclusion}.

\section{Related Works} \label{Related}

Recent works have leveraged self-supervised features to enhance the performance of audio deepfake detection. To improve generalization against unseen attacks, several methods have adopted strategies such as one-class learning, knowledge distillation, and latent space optimization. Lu \emph{et al.} \cite{10446270} proposed a one-class knowledge distillation approach, where a teacher model is trained on both bona fide and spoofed speech, while the student model only has access to bona fide data and learns to mimic the teacher’s output, thereby capturing the bona fide distribution. Kim \emph{et al.} \cite{kim24b_interspeech} proposed an adaptive centroid shift (ACS) method that continually updates the centroid of bona fide representations, enhancing one-class learning by clustering real samples into a compact space that is robust to spoofing. Wang \emph{et al.} \cite{FTDKD} proposed FTDKD, which uses frequency and time-domain knowledge distillation to transfer information from a teacher model trained on high-quality data to a student model trained on low-quality, compressed data to enhance the detection performance of low-quality compressed audio. Zhu \emph{et al.} \cite{SLIM} proposed the Style-Linguistics Mismatch (SLIM) method, which leverages self-supervised learning on only bona fide samples to learn style-linguistic dependencies, later combined with pretrained acoustic features to separate deepfake from bona fide samples. Huang \emph{et al.} \cite{LSR+LSA} proposed Latent Space Refinement (LSR) and Latent Space Augmentation (LSA), introducing learnable prototypes for the spoof class and applying latent space augmentations to improve generalization.

Several methods employ various feature fusion strategies, including multi-layer, multi-scale, and cross-modal integration, to enhance the discriminative capability of self-supervised features. Pan \emph{et al.} \cite{attentive_merge} proposed an attentive merging method to aggregate hierarchical hidden embeddings from the WavLM model for anti-spoofing. Zhang \emph{et al.} \cite{SLS} utilized Sensitive Layer Selection (SLS), which captures sensitive contextual information from layer-wise XLS-R features for fake audio detection. Truong \emph{et al.} \cite{TCM} proposed the Temporal-Channel Modeling (TCM) module to enhance multi-head self-attention based on the head tokens designed to capture temporal-channel dependency of the input sequence. Jin \emph{et al.} \cite{10890563} proposed WaveSpec, which adopts wav2vec to extract waveform features and U-Net to extract multi-scale spatial features, then uses Cross-Modal Feature Fusion (CMFF) to amplify the advantages of different modal information. Wang \emph{et al.} \cite{MOE} proposed a feature fusion strategy based on the Mixture of Experts (MoE) framework, guided by a gating network that adaptively selects relevant features across different layers. Guo \emph{et al.} \cite{guo2024audio} proposed a Multi-Fusion Attentive (MFA) classifier that integrates information across time and feature layers using attentive statistics pooling. Tran \emph{et al.} \cite{tran24_interspeech} proposed using layer-wise attentive statistics pooling on WavLM to enhance the discriminative power of utterance-level features.

\begin{figure*}[t]
\centering
\includegraphics[width=\textwidth]{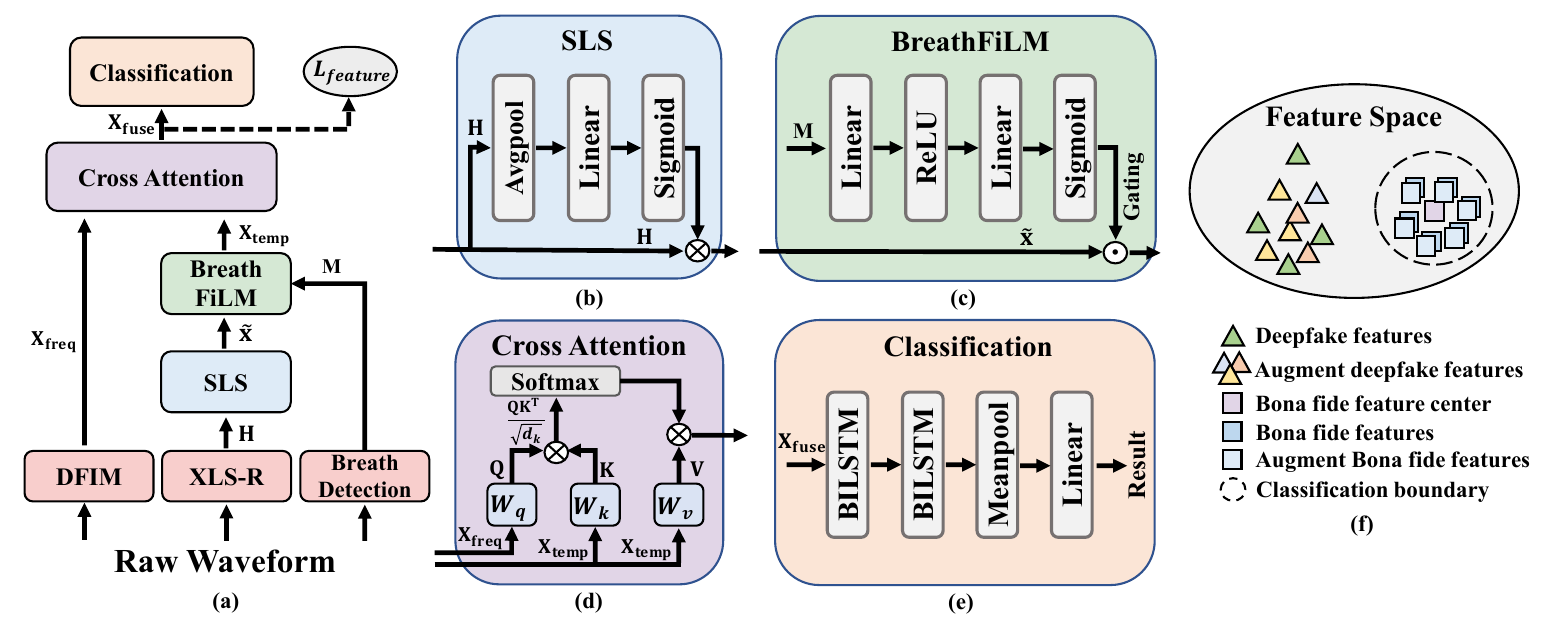}
\caption{An overview of BreathNet.}
\label{fig-pipeline}
\end{figure*}

Several methods propose architectural modifications that aim to leverage self-supervised features more effectively. Tak \emph{et al.} \cite{hemlata_wav2vec2} first integrated wav2vec 2.0 features into the spectro-temporal graph attention network AASIST, demonstrating the benefits of combining self-supervised learning with graph-based modeling. Zhang \emph{et al.} \cite{10448049} proposed AASIST2, enhancing the original AASIST architecture by replacing standard residual blocks with Res2Net blocks to enable multi-scale feature extraction. Xiao \emph{et al.} \cite{Mamba} proposed XLSR-Mamba, which integrates wav2vec 2.0 with a dual-column bidirectional Mamba architecture. In this design, two parallel columns process forward and backward features, and are subsequently fused to capture both local and global dependencies. Liu \emph{et al.} proposed Nes2Net \cite{liu2025nes2net}, a Res2Net-based architecture that retains the expressive power of high-dimensional features while reducing overall model complexity.

Layton \emph{et al.}~\cite{10.1145/3754456} proposed a breath detection framework in which a breath detector is used to explicitly identify breath events and derive global breathing statistics, such as breaths per minute, average breath duration, and average spacing between breaths. These aggregated statistics are then employed as primary decision features for deepfake detection. In contrast, our framework does not rely on explicit breath metrics for classification. Instead, we design an auxiliary module that leverages the breath mask to guide the XLS-R feature extractor to learn breath-related cues during training.

\section{Methodology} \label{Methodology}

As shown in Fig. \ref{fig-pipeline}, our method follows a dual-branch architecture that extracts information from both the temporal and frequency branches. The temporal branch leverages a pretrained XLS-R model with an SLS and BreathFiLM module, while the frequency branch employs a DFIM module to capture complementary frequency-domain information. The two branches’ features are fused via a cross-attention that adaptively aligns their representations. The fused features are then fed into a classifier for prediction. To further enhance discriminability, we introduce an additional feature loss that encourages better separation between bona fide and deepfake samples in the feature space. 

\subsection{Temporal Branch}

\subsubsection{XLS-R feature encoder}
We adopt the self-supervised, cross-lingual speech representation model XLS-R \cite{babu22_interspeech} as our feature extraction. XLS-R combines CNN and Transformer layers to process raw audio, where the CNN downsamples the waveform into latent features, and the Transformer captures contextual representations. In our framework, we directly extract features from raw waveforms using the pre-trained XLS-R model and fine-tune it jointly with our audio deepfake detection network. This end-to-end training improves the model's ability to distinguish bona fide and deepfake speech.

\subsubsection{Sensitive Layer Select Module}

We adopt the Sensitive Layer Select (SLS) module~\cite{SLS} to adaptively aggregate hidden output from all transformer layers in the XLS-R model. Given the layer-wise outputs $\mathbf{H} = [\mathbf{h}_{1}, \mathbf{h}_{2}, \dots, \mathbf{h}_{L}] \;\in\;\mathbb{R}^{L \times T \times D}$, where $L$ is the total number of layers, $T$ is the temporal length, and $D$ is the feature dimensionality. SLS first applies adaptive average pooling over the temporal axis:
\begin{equation}
\mathbf{s} = \mathrm{avgpool}_{\text{T}}(\mathbf{H}).
\end{equation}

The pooled feature $\mathbf{s} \in \mathbb{R}^{L \times D}$ is then passed through a linear layer, followed by a sigmoid activation \(\sigma(\cdot)\) to produce normalized layer-importance weights:
\begin{equation}
\mathbf{w} = \sigma\bigl(\mathrm{FC}(\mathbf{s})\bigr).
\end{equation}

$\mathbf{w} = [w_1, w_2, \dots, w_L] \in \mathbb{R}^L$ contains the learned importance weights for each of the transformer layers, where $w_i$ denotes the importance of the $i$-th layer. These scores are used to perform a weighted summation across all layers:
\begin{equation}
\mathbf{\widetilde{x}} = \sum_{i=1}^{L} w_{i} \cdot \mathbf{h}_{i},
\end{equation}

The SLS module performs a learnable aggregation over all transformer layers by assigning adaptive weights based on the relative importance of each layer. This allows the model to emphasize the most informative representations for audio deepfake detection.

\subsubsection{BreathFiLM Module}

Breath sounds are a subtle yet essential component of natural speech perception. However, they are often inaccurately modeled in synthetic speech, as many speech synthesis systems tend to overlook this crucial element. Consequently, these regions are typically poorly synthesized and can serve as reliable cues for detecting audio deepfakes. Motivated by this, we propose BreathFiLM (Breath-based Feature-wise Linear Modulation), a modulation mechanism that adaptively learns to emphasize breath-related frames based on a binary breath mask. BreathFiLM is trained jointly with the XLS-R backbone, guiding the feature extractor to learn breath-related cues and to incorporate them into the temporal features. As a result, the model can better exploit breath-related cues, thereby improving its discriminability for deepfake audio detection.

Specifically, we utilize a self-training breath detection model \cite{yang24c_interspeech} to distinguish between breath and non-breath segments in each audio sample. This model utilizes a Conformer-based encoder to capture both local and global dependencies, complemented by a bidirectional LSTM decoder that generates temporally aligned predictions with fine-grained resolution. In the original work, the breath detector was trained on the LibriTTS-R \cite{koizumi23_interspeech} corpus and achieved an IoU of 0.836, precision above 0.92, and recall around 0.89 on the test set, demonstrating strong breath detection performance. However, the model outputs breath intervals as continuous time segments rather than frame-level labels. Since our modulation mechanism requires a frame-wise binary mask aligned with the temporal feature structure, we introduce a preprocessing procedure to bridge this mismatch. All input audio is first normalized to a fixed duration of $n$ seconds to match the input requirements of the detection model. Longer clips are truncated, while shorter ones are extended by repeating their content. The normalized waveform is then divided into $T$ non-overlapping frames, and the detected breath intervals are mapped to a binary mask over these frames:
\begin{equation}
\mathbf{M} = f_{\mathrm{breath}}(\mathbf{H}) = [\,m_{1}, m_{2}, \dots, m_{T}\,]\;\in\;\{0,1\}^{T},
\end{equation}
where $f_{\mathrm{breath}}(\cdot)$ denotes the breath detection model. Each binary value $m_{t}=1$ indicates that a breath sound is detected in frame $t$, while $m_{t}=0$ denotes its absence.

After obtaining the binary breath mask for each sample, we employ a feature-wise linear modulation mechanism implemented with a lightweight multilayer perceptron (MLP). This module enables fine-grained, frame-level adjustment of temporal features based on the presence or absence of breath. As shown in Figure~\ref{fig-pipeline}, part (C) illustrates the stage where the binary mask $ \mathbf{M}$ is first passed through a two-layer MLP. The first linear layer projects the mask to a hidden representation, followed by a ReLU activation. The second layer maps this representation to the temporal feature dimension $D$, followed by a sigmoid activation to obtain the gating values:

\begin{equation}
\mathbf{g} = \sigma\bigl(\mathbf{W_{2}}\,\mathrm{ReLU}(\mathbf{W_{1}}\,\mathbf{M})\bigr)\;,
\end{equation}
where \(\mathbf{W_{1}}\in\mathbb{R}^{1\times H}\), \(\mathbf{W_{2}}\in\mathbb{R}^{H\times D}\) are the weights of the MLP, and \(\sigma(\cdot)\) represents the sigmoid activation function. 

The gating values are designed to selectively highlight regions related to breathing. To preserve information from non-breath frames while enhancing breath-related ones, we adopt a residual gating strategy. Specifically, we add a constant bias of 1 to the learned gating values so that non-breath frames are preserved, while breath frames receive additional emphasis. These gating weights are then applied to the temporal features via element-wise multiplication, yielding the modulated time-domain representation $\mathbf{x_{temp}}$.

\begin{equation}
\mathbf{G} = 1 + \mathbf{g}, \; \mathbf{x_{temp}} = \mathbf{G} \odot \mathbf{\widetilde{x}}.
\end{equation}

The core intuition behind BreathFiLM is that breath sounds carry meaningful cues for distinguishing synthetic speech. A simple binary-mask-driven modulation is sufficient to effectively guide the network to focus on breath regions.

\subsection{Frequency Branch}

While wav2vec-based features have achieved strong results in various speech tasks, they are primarily derived from time-domain waveforms. In addition to temporal information, frequency-domain features can also be leveraged for audio deepfake detection.

In speech signal processing, a common approach to convert waveforms from the time domain to the frequency domain is through the Fourier transform. However, sinc convolution (SincConv) \cite{sinc} is a more interpretable alternative to standard convolution operations. Instead of learning unconstrained filter weights, SincConv parameterize each filter as a band-pass filter defined by its low and high cutoff frequencies. This design enables neural networks to learn frequency-domain filters in an end-to-end manner. In this work, we employ a SincConv-based feature, Discriminative Frequency Information Mining (DFIM) \cite{DFIM}, to extract discriminative frequency information from raw audio waveforms directly. DFIM first applies pre-emphasis to enhance high-frequency components, followed by a SincConv layer consisting of multiple parameterized band-pass filters. This operation produces a time-frequency representation, which is subsequently passed through adaptive max pooling, batch normalization, and SELU.

We believe that the frequency-domain branch can provide additional discriminative information that complements temporal features. To enable effective integration with temporal features, we apply a projection layer to the DFIM output, yielding the final frequency-domain feature $\mathbf{x_{freq}}$ that is dimensionally aligned with the temporal feature $D$.

\subsection{Feature Refinement}

\subsubsection{Feature Fusion}

Cross-attention \cite{chen2021crossvit} is employed to enable one feature to selectively attend to another, making it particularly effective for multi-source feature fusion tasks. Unlike self-attention, which focuses on relationships within a single feature, cross-attention builds connections between two different feature sequences. In our case, we adopt a cross-attention mechanism where frequency features serve as queries and temporal features act as keys and values. This enables the model to dynamically merge spectral and temporal information, yielding a fused representation $\mathbf{X_{fuse}}$ that enhances its ability to detect deepfake audio.

Given $\mathbf{X_{freq}}$ and $\mathbf{X_{temp}}$, we project them into separate subspaces using trainable linear transformations:
\begin{equation}
\mathbf{Q} = \mathbf{X_{freq}} \mathbf{W^Q},\quad \mathbf{K} = \mathbf{X_{temp}} \mathbf{W^K},\quad \mathbf{V} = \mathbf{X_{temp}} \mathbf{W^V},
\end{equation}
where $\mathbf{W^Q} \in \mathbb{R}^{D \times d_q}$ ,$\mathbf{W^K} \in \mathbb{R}^{D \times d_k}$, $\mathbf{W^V} \in \mathbb{R}^{D \times d_v}$ are learnable parameters.

The attention mechanism computes relevance scores using scaled dot-product attention:
\begin{equation}
\text{Attention}(\mathbf{Q, K, V}) = \text{softmax}\left(\frac{\mathbf{QK^\top}}{\sqrt{d_k}}\right)\mathbf{V},
\end{equation}
where $d_k$ is the dimension of the key. 

To enhance the model’s capacity to capture diverse dependencies, multi-head attention splits the input into \( h \) heads and computes parallel attention across each head. In this work, we employ $h=8, d_q=d_k=d_v=1024$, and no dropout is applied. This fusion enables the model to effectively integrate complementary discriminative cues from the frequency domain to temporal features, improving detection ability for deepfakes.

\subsubsection{Refinement Loss}

To enhance the model’s ability to distinguish deepfakes from bona fide samples, we apply a feature-level loss on the fused features. Specifically, we employ the Positive-only Supervised Contrastive Loss (PSCL) and center loss to reinforce the compactness of the bona fide class, while the contrast loss is utilized to enforce inter-class separability. Each loss term plays a distinct and complementary role. The ideal feature distribution is illustrated in Fig.~\ref{fig-pipeline}(f).

\textbf{Positive-only Supervised Contrastive Loss.}
Supervised contrastive learning (SCL) \cite{khosla2020supervised} extends conventional contrastive learning by utilizing label information to define positive and negative sample pairs. It encourages embeddings from the same class to be pulled closer together in the feature space while pushing embeddings of different classes apart. 

However, in our training set, deepfake samples significantly outnumber bona fide ones, leading to a severe class imbalance. Under the SCL setting, imbalance will bias the optimization toward the deepfake class and consequently reduce the intra-class compactness of bona fide representations. To address this issue, we design the PSCL, which exclusively leverages bona fide samples as anchors by discarding the inter-class separability objective of SCL, thereby focusing solely on ensuring that bona fide samples exhibit sufficient similarity in the feature space. 

In PSCL, we enhance the diversity of bona fide samples by injecting scaled Gaussian noise, which increases intra-class variability and improves the model’s generalization. This strategy ensures sufficient variation from the bona fide class, which is crucial for learning compact representations. Let $\mathbf{z}$ denote the feature representation of a sample. For each bona fide feature $\mathbf{z}_i^{\text{bona}}$, we generate the augmented feature by sampling noise $\boldsymbol{\epsilon} \sim \mathcal{N}(0, 1)$:

\begin{equation}
\mathbf{z}_i^{\text{bona+}}=\mathbf{z}_i^\text{bona}+\delta \cdot \epsilon,
\end{equation}
where $\delta$ is a scalar that controls the intensity of the noise.

The augmented features are used as additional positive samples, jointly optimized with their original bona fide in the contrastive loss. Let $D_{bona}$ denote the set of bona fide samples, including original features $\mathbf{z}_i^{\text{bona}}$ and their augmented variants $\mathbf{z}_i^{\text{bona+}}$. The PSCL is defined as:

\begin{align}
\mathcal{L}_{\text{PSCL}} = 
& -\frac{1}{|D_{bona}|} \sum_{i \in D_{bona}} \frac{1}{|D_{bona}|-1} \sum_{j \in D_{bona} \setminus \{i\}} \notag \\
& \log \frac{\exp \left( s(\mathbf{z}_i, \mathbf{z}_j)/\tau \right)}
{\sum\limits_{k \in D_{bona} \setminus \{i\}} \exp \left( s(\mathbf{z}_i, \mathbf{z}_k)/\tau \right)},
\end{align}
where $s(\mathbf{z}_i, \mathbf{z}_j)$ denotes the cosine similarity between feature $\mathbf{z}_i$ and $\mathbf{z}_j$, and $\tau$ is a temperature parameter.

\textbf{Center Loss.}
While the PSCL achieves intra-class compactness by enhancing the similarity among bona fide features, it operates at a relatively coarse level. Therefore, we introduce a fine-grained constraint to further refine compactness within the bona fide class. Specifically, we adopt a center loss that pulls all bona fide features toward a learned class center in the embedding space. The center loss is as follows:

\begin{equation}
\mathcal{L}_{\text{center}} = \frac{1}{|D_{bona}|} \sum_{i \in D_{bona}} \frac{1 - s(\mathbf{z}_i, \mathbf{c})}{2},
\end{equation}
where $\mathbf{c}$ is the bona fide center.

The bona fide center $\mathbf{c}$ is initialized as the mean of bona fide features in the first batch. For the subsequent batch, we compute the mean of the bona fide features $\mathbf{\bar{z}}$ and update the center using a momentum-based strategy:

\begin{equation}
    \mathbf{c} \leftarrow \mu \cdot \mathbf{c}+(1-\mu) \cdot \mathbf{\bar{z}},
\end{equation}
where $\mu$ is a momentum coefficient.

The center loss encourages bona fide samples to form tighter clusters in the feature space, thereby reducing intra-class variation. As a result, the model learns more compact and well-structured feature representations. However, this center-based constraint is applied exclusively to the bona fide class, as deepfake samples are typically generated using diverse synthesis methods, making it difficult to form a reliable center.

\textbf{Contrast Loss.}
Deepfake samples exhibit substantial intra-class variability, making it impractical to establish a meaningful center. Therefore, instead of pulling them toward an unreliable cluster, we impose a constraint that explicitly pushes deepfake features away from the bona fide center, effectively improving class separability in the feature space. To simulate unseen synthesis methods that enhance the model’s generalization to diverse deepfakes, we employ a targeted augmentation strategy based on mixup. Specifically, we generate new deepfake samples by mixing pairs of deepfake features within each batch, creating new feature representations that mimic potential but unseen forgeries. The contrast loss is designed to push deepfake samples away from the bona fide center, thereby providing a complementary effect to the PSCL and center loss. Let $D_{fake}$ denote the set of deepfake samples from the training set. The contrast loss is defined as:
\begin{align}
\mathcal{L}_{\text{contrast}} =\ 
& \frac{1}{|D_{fake}|} \sum_{l \in D_{fake}} \frac{1 + s(\mathbf{z}_l, \mathbf{c})}{2} \nonumber \\
+ & \frac{1}{\binom{|D_{fake}|}{2}} \sum_{1 \leq n < m \leq |D_{fake}|} \frac{1 + s\left({\frac{\mathbf{z}_n + \mathbf{z}_m}{2}, \mathbf{c}}\right)}{2}.
\end{align}

By mixing every pair of deepfake samples, we enhanced the diversity of unseen variations. As a result, the contrastive loss further increases the difference between deepfake and bona fide samples in the feature space, thereby enabling the model to learn more robust decision boundaries and improving discriminative performance.

\textbf{Feature Loss.}
Our feature loss comprises three components. PSCL encourages bona fide features with sufficient similarity, while the center and contrast losses offer fine-grained constraints: the center loss pulls bona fide features toward a compact cluster, and the contrast loss pushes deepfake features away from the bona fide center. This joint design enhances feature separability, leading to more discriminative representations and ultimately improving the model’s generalization ability across diverse datasets. The feature loss is defined as follows:
\begin{equation}
\mathcal{L}_{\text{feature}} = \mathcal{L}_{\text{PSCL}} + \alpha \mathcal{L}_{\text{center}} + \beta \mathcal{L}_{\text{contrast}},
\end{equation}
where \( \alpha \) and \( \beta \) are hyperparameters to balance the contribution of each loss term.

\begin{table}[t]
\centering
\caption{Feature shapes at different stages. $N$ represents the batch size.}
\label{tab:shapes}
% Reduce column separation and font size to fit
\setlength{\tabcolsep}{1.2mm}
{
\begin{tabular}{lll}
\hline
\toprule
\textbf{Stage} & \textbf{Module}                 & \textbf{Shape}             \\
\midrule
Input (Raw waveform) & –                 & $N\times64600$             \\
\midrule
\multirow{5}{*}{Freq. Feature} &
  Pre-Emphasis        &                            \\
& SincConv            &                            \\
& AdaptiveMaxPool2d   &                            \\
& BatchNorm2d \& SELU &                            \\
& Projection          & \multirow{-5}{*}{$N\times32\times1024$} \\
\midrule
\multirow{3}{*}{Temp. Feature} &
  XLS-R               &                            \\
& SLS                 &                            \\
& BreathFiLM          & \multirow{-3}{*}{$N\times201\times1024$} \\
\midrule
Fusion Feature & Cross Attention       & $N\times32\times1024$      \\
\midrule
\multirow{4}{*}{Classification} &
  BiLSTM1             &                            \\
& BiLSTM2             &                            \\
& Mean Pool           &                            \\
& Linear              & \multirow{-4}{*}{$N\times2$}             \\
\bottomrule
\hline
\end{tabular}
}
\end{table}

\subsection{Classifier}

We employ a BiLSTM network and linear layers to map the learned representations to the label space. BiLSTM is a type of recurrent neural network that extends the traditional LSTM by processing sequences in both forward and backward directions. This allows it to capture more complete contextual information from the input. Since speech is inherently sequential, deepfake audio often exhibits subtle irregularities or artifacts in temporal transitions. By modeling the sequence bidirectionally, BiLSTM is better equipped to detect such anomalies. 

To optimize the final classification layer under class imbalance, we adopt the weighted CrossEntropy loss. Given the predicted class probabilities $\hat{y}$ and the one-hot ground-truth label $y$ , the loss is computed as:
\begin{equation}
\mathcal{L}_{\text{softmax}} = - \sum_{i=1}^{N_{c}} w_i  y_i \log(\hat{y}_i),
\label{eq}
\end{equation}
where $N_{c}$ is the number of classes, $w_{i}$ is the weight assigned to each class.

The total loss consists of the classification loss and the feature loss:
\begin{equation}
\mathcal{L}_{\text{total}} = \mathcal{L}_{\text{softmax}} + \lambda \mathcal{L}_{\text{feature}},
\end{equation}
where $\lambda$ is a hyperparameter used to balance these two losses.

As shown in Table~\ref{tab:shapes}, we illustrate each module in our proposed method along with the corresponding input shape transformations. For different modules, we employ a different learning rate during training. Specifically, the XLS-R backbone is fine-tuned with a smaller learning rate to preserve its pre-trained representations, while the remaining trainable modules are updated with relatively larger learning rates. This strategy helps maintain stable feature extraction in the early layers while enabling effective adaptation of the task-specific components.

\section{Experiment Results} \label{Results}

\subsection{Experimental Setups}

\subsubsection{Datasets and Metrics}
We conduct experiments on five publicly available datasets: ASVspoof 2019 LA (19LA) \cite{wang2020asvspoof}, ASVspoof 2021 LA (21LA), ASVspoof 2021 DF (21DF) \cite{liu2023asvspoof}, In-the-Wild (ITW) \cite{muller22_interspeech}, and ASVspoof5 \cite{WANG2026101825}. The 19LA contains spoofed speech generated using 19 different VC and TTS systems. Among them, 6 systems are included in the training sets, while the remaining 13 systems appear only in the evaluation set. The 21LA is derived from the 19LA evaluation set, where utterances are processed with multiple lossy codecs and transmission simulators to undergo transmission across either a VoIP or PSTN+VoIP network. The 21DF focuses exclusively on high-fidelity deepfake attacks, synthesized by 100 different spoofing attack algorithms. To emulate real-world usage, the data is encoded with various lossy codecs commonly employed in media storage and then decoded to recover uncompressed audio. The ITW consists of audio samples from 58 celebrities and politicians. Both bona fide and deepfake speech samples are collected from publicly available sources, such as social networks and video streaming platforms, which may contain background noise. The ASVspoof5 dataset is designed with a strong emphasis on real-world conditions and generalization to unseen spoofing attacks. It adopts strict attack-disjoint splits across the training, development, and evaluation sets, which contain 8, 8, and 16 distinct spoofing attack algorithms, respectively, ensuring that all evaluation-time attacks are unseen during training. Moreover, ASVspoof5 is the first dataset in the ASVspoof series to systematically incorporate adversarial attacks, which are designed to mislead detection models through targeted perturbations rather than introducing overt synthesis artifacts, further increasing the realism and difficulty of the task. The dataset details are in Table \ref{tab:dataset_info}.

Now, the prevailing evaluation protocol involves training the model on the 19LA training set, followed by a comprehensive assessment on the 19LA, 21LA, 21DF, and ITW evaluation sets to evaluate the model’s capability. In contrast, ASVspoof5 adopts a self-contained evaluation protocol, where models are trained and evaluated solely within the ASVspoof5 dataset using its predefined training, development, and evaluation partitions. We evaluate model performance using the equal error rate (EER), the standard metric in the ASVspoof Challenges. In addition, following the ASVspoof5 evaluation protocol, we also report the cost of log-likelihood ratios (CLLR) and the minimum detection cost function (minDCF) to provide a more comprehensive assessment of detection performance. 

EER indicates the inherent separability between bona fide and spoofed speech based on detection scores. minDCF evaluates the minimum achievable detection cost under predefined error costs and class priors, assuming an oracle decision threshold determined with access to ground-truth labels. CLLR is used to assess the quality and calibration of detection scores interpreted as log-likelihood ratios. Lower EER and minDCF indicate better detection performance, while a lower CLLR reflects better score quality and calibration.

\subsubsection{Baselines}
We conduct a comprehensive comparison of audio deepfake detection methods published over the past two years, with a particular focus on approaches that utilize self-supervised front-end feature extractors. The baselines include Wav2Vec + AASIST \cite{hemlata_wav2vec2}, MFA \cite{guo2024audio}, AASIST2 \cite{10448049}, OCKD \cite{10446270}, AttM-LSTM \cite{attentive_merge}, ASP + MLP \cite{tran24_interspeech}, OC-ACS \cite{kim24b_interspeech}, TCM \cite{TCM}, SLS \cite{SLS}, FTDKD \cite{FTDKD}, SLIM \cite{SLIM}, WaveSpec \cite{10890563}, MoE \cite{MOE}, LSR + LSA \cite{LSR+LSA}, XLSR-Mamba \cite{Mamba} and Nes2Net-X \cite{liu2025nes2net}. We adopt the best reported results of each method.

% To ensure fairness, we adopt the best reported results of each method, without additional codec augmentation, for comparison. Moreover, all results are based on a single model across all eval sets, rather than using different models for different eval sets.

\begin{table}[t]
\centering
\caption{Detailed information of the train and eval datasets used in our experiments.}
\label{tab:dataset_info}
\setlength{\tabcolsep}{3mm}
{
\begin{tabular}{lrrr}
\hline
\toprule
\multirow{2}{*}{Set}      & \multicolumn{1}{c}{Bonafide} & \multicolumn{1}{c}{Spoofed} & \multicolumn{1}{c}{Total} \\ 
                          & \multicolumn{1}{c}{\#utterance}   & \multicolumn{1}{c}{\#utterance}   & \multicolumn{1}{c}{\#utterance}   \\ 
\midrule
Train (2019 LA)           &   2{,}580               &   22{,}800            &   25{,}380            \\
Eval (2019 LA)            &   7{,}355               &   63{,}882            &   71{,}237            \\
Eval (2021 LA)            &  14{,}816               &  133{,}360            &  148{,}176            \\
Eval (2021 DF)            &  14{,}869               &  519{,}059            &  533{,}928            \\
Eval (In-the-Wild)        &  19{,}963               &   11{,}816            &   31{,}779            \\
\midrule
Train (ASVspoof5)       &   18{,}797               &   16{,}3560            &   18{,}2357           \\
Dev (ASVspoof5)         &   31{,}334               &   10{,}9616            &   14{,}0950           \\
Eval (ASVspoof5)        &   13{,}8688               &   54{,}2086            &   680{,}774            \\
\bottomrule
\hline
\end{tabular}
}
\end{table}

\subsection{Implementation Details}

All audio data are cropped or concatenated to create segments of around 4 seconds duration. Temporal features are extracted directly from raw waveforms using a pre-trained XLS-R model (0.3B parameters) implemented in the Fairseq framework. For 19LA training, the XLS-R encoder and the back-end classifier are optimized jointly with the Adam optimizer, where the initial learning rate and weight decay are set to 0.00001 and 0.0001, respectively. The learning rate for the XLS-R encoder is set to 0.000001. Training is conducted with a batch size of 10 for up to 50 epochs, with early stopping applied if the training loss does not improve for two consecutive epochs. The final evaluation model is obtained by averaging the weights of the last three checkpoints. The class weights for the weighted cross-entropy loss are set to 0.9 for bona fide and 0.1 for deepfake samples. Data augmentation is performed using the RawBoost algorithm (4) \cite{9746213} to introduce convolutive noise, impulsive signal-dependent additive noise, and stationary signal-independent additive noise. For ASVspoof5 training, our models are trained using settings similar to those adopted for Nes2Net \cite{liu2025nes2net} on the ASVspoof5 dataset, except that the batch size is set to 10. For the feature loss, the hyperparameters are set to $\lambda=0.5$, $\alpha=1$, $\beta=1$, $\mu=0.9$, and $\delta=0.1$. All experiments are conducted on a single GeForce RTX 3090 GPU.

\begin{figure*}[t]
\centering
\includegraphics[width=\textwidth]{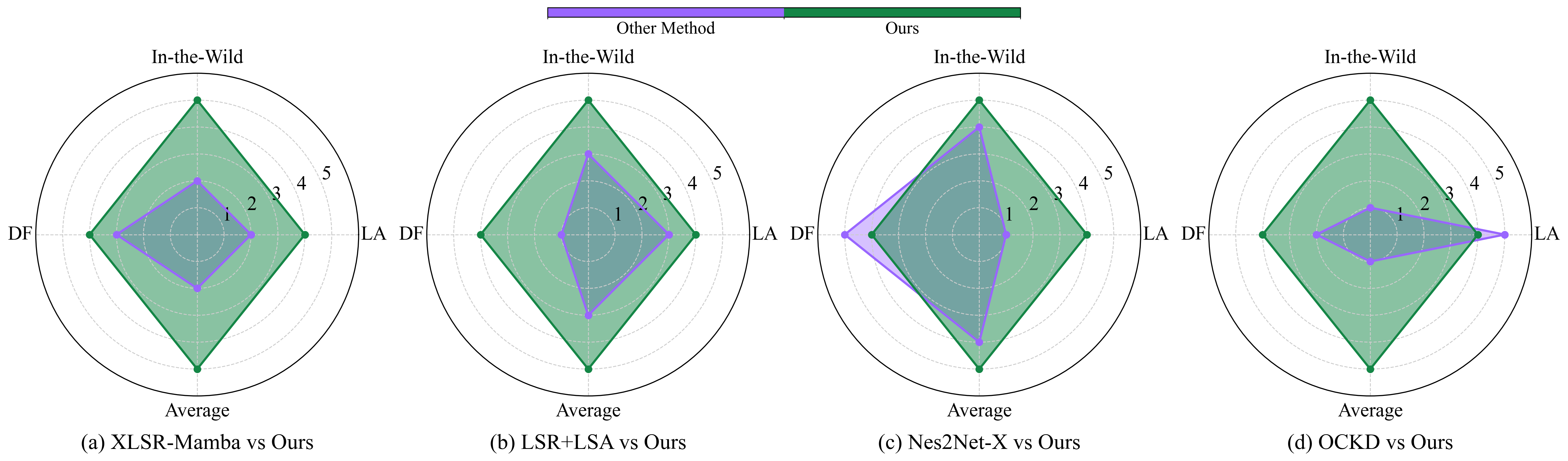}
\caption{
The performance comparison of our method with four baselines. A larger enclosed area indicates better performance.
}
\label{fig-radar}
\end{figure*}

\begin{table}[t]
\centering
\caption{Performance in EER (\%) on the ASVspoof 2019.}
\setlength{\tabcolsep}{5mm}
{
\begin{tabular}{llcc}
\hline
\toprule
Publication & System & 19LA \\
\hline
\midrule

ICASSP 2024 & MFA~\cite{guo2024audio}                   & 0.42 \\ 
ICASSP 2024 & AASIST2~\cite{10448049}                   & \textbf{0.15} \\
ICASSP 2024 & OCKD~\cite{10446270}                      & 0.39 \\
Interspeech 2024 & AttM-LSTM~\cite{attentive_merge}     & 0.65 \\ 
Interspeech 2024 & ASP+MLP~\cite{tran24_interspeech}    & 0.23 \\  
Interspeech 2024 & OC-ACS~\cite{kim24b_interspeech}     & 0.17 \\
TASLP 2024 & FTDKD~\cite{FTDKD}                         & 0.30 \\
NeurIPS 2024 & SLIM~\cite{SLIM}                         & 0.20 \\
ICASSP 2025 & MoE~\cite{MOE}                            & 0.74 \\
ICASSP 2025 & LSR+LSA~\cite{LSR+LSA}                    & \textbf{0.15} \\

\midrule

\rowcolor{gray!30} Ours & Ours   & 0.23 \\

\bottomrule
\hline
\end{tabular}
}
\label{tab_asvspoof19}
\end{table}

\begin{table}[t]
\centering
\caption{Performance in EER (\%) on the ASVspoof 2021 LA and DF.}
{
\begin{tabular}{llccc}
\toprule
\multirow{2}{*}{Publication} & \multirow{2}{*}{System} 
& \multicolumn{3}{c}{ASVspoof 2021} \\ 
\cmidrule(r){3-5}
 &  & LA & DF & Avg. \\
\midrule 

Odyssey 2022 
& Wav2vec+AASIST~\cite{hemlata_wav2vec2}~${\dagger}$    
& \textbf{0.82} & 2.85 & 1.84 \\ 

ICASSP 2024 
& MFA~\cite{guo2024audio}                     
& 5.08 & 2.56 & 3.82 \\

ICASSP 2024 
& AASIST2~\cite{10448049}                     
& 1.61 & 2.77 & 2.19 \\

ICASSP 2024 
& OCKD~\cite{10446270}                        
& \underline{0.90} & 2.27 & 1.59 \\

Interspeech 2024 
& AttM-LSTM~\cite{attentive_merge}       
& 3.50 & 3.19 & 3.35 \\  

Interspeech 2024 
& ASP+MLP~\cite{tran24_interspeech}      
& 3.31 & 4.47 & 3.89 \\   

Interspeech 2024 
& OC-ACS~\cite{kim24b_interspeech}       
& 1.30 & 2.19 & 1.75 \\

Interspeech 2024 
& TCM~\cite{TCM}~${\dagger}$                         
& 1.03 & 2.06 & \underline{1.55} \\

ACMMM 2024 
& SLS~\cite{SLS}                               
& 2.87 & 1.92 & 2.40  \\

TASLP 2024 
& FTDKD~\cite{FTDKD}                           
& 2.96 & 2.82 & 2.89   \\

NeurIPS 2024 
& SLIM~\cite{SLIM}                           
& -- & 4.40 & -- \\

ICASSP 2025 
& WaveSpec~\cite{10890563}                    
& -- & 1.90 & -- \\

ICASSP 2025 
& MoE~\cite{MOE}                              
& 2.96 & 2.54 & 2.75 \\

ICASSP 2025 
& LSR+LSA~\cite{LSR+LSA}                      
& 1.19 & 2.43 & 1.81 \\

SPL 2025 
& XLSR-Mamba~\cite{Mamba}~${\dagger}$                       
& 1.68 & 1.88 & 1.78 \\

TIFS 2025 
& Nes2Net-X~\cite{liu2025nes2net}               
& 1.88 & \textbf{1.49} & 1.69 \\

\midrule
\rowcolor{gray!30} 
Ours 
& Ours  
& 1.15 & \underline{1.87} & \textbf{1.51} \\

\bottomrule
\end{tabular}
}

\vspace{1mm}
\begin{minipage}{\linewidth}
\footnotesize\raggedright
${\dagger}$ LA and DF results are obtained using two different rawboost augmentation configurations.
\end{minipage}
\label{tab_asvspoof21}
\end{table}

\begin{table}[t]
\centering
\caption{Performance in EER (\%) on the In-the-Wild.}
\setlength{\tabcolsep}{5mm}
{
\begin{tabular}{llcc}
\hline
\toprule
Publication & System & ITW \\
\hline
\midrule
ICASSP 2024 & OCKD~\cite{10446270}            & 7.68 \\
ACMMM 2024 & SLS~\cite{SLS}                   & 7.46 \\
NeurIPS 2024 & SLIM~\cite{SLIM}               & 12.50 \\
ICASSP 2025 & WaveSpec~\cite{10890563}        & 6.58 \\
ICASSP 2025 & MoE~\cite{MOE}                  & 12.48 \\
ICASSP 2025 & LSR+LSA~\cite{LSR+LSA}          & 5.92 \\
SPL 2025 & XLSR-Mamba~\cite{Mamba}            & 6.71 \\
TIFS 2025 & Nes2Net-X~\cite{liu2025nes2net}   & 5.52 \\
\midrule

\rowcolor{gray!30} Ours & Ours    & \textbf{4.70} \\

\bottomrule
\hline
\end{tabular}
}
\label{tab_inthewild}
\end{table}

\begin{table}[t]
\caption{Comparison of the proposed method on the ASVspoof5 dataset. Baseline results are reported by~\cite{liu2025nes2net}}
\label{tab:asv5}
\centering
\setlength{\tabcolsep}{5mm}
{
\begin{tabular}{cccccc}
\hline
\toprule
Model & CLLR & minDCF & EER \\
\hline
\midrule
AASIST    & 0.9587 & 0.1645 & 6.08 \\
Nes2Net   & 0.7912 & 0.1568 & 6.13 \\
Nes2Net-X & 0.7344 & 0.1535 & 5.92 \\
\midrule
\rowcolor{gray!30} Ours      & \textbf{0.7134} & \textbf{0.1307} & \textbf{4.94} \\
\bottomrule
\hline
\end{tabular}
}
\end{table}

\subsection{Main Results}

As shown in Table~\ref{tab_asvspoof19}, our method achieves an EER of 0.23\% on 19LA, which is comparable to the state of the art. Table~\ref{tab_asvspoof21} reports the results on 21LA and DF. The 21LA simulates telecommunication transmission conditions, with speech encoded using various telephony codecs and transmitted through VoIP or PSTN systems, thereby emphasizing robustness to channel and codec distortions. In contrast, the 21DF is designed to reflect more realistic deepfake scenarios, where a broader range of spoofing generation methods is considered and most utterances undergo lossy compression, resulting in unpredictable acoustic variations.

Given that the two tracks focus on substantially different characteristics, optimizing a single model to perform well on both tracks is inherently challenging. Our model achieves EERs of 1.15\% and 1.87\% on 21LA and 21DF, respectively, demonstrating competitive performance. When jointly considering the results across the two different tracks, our method attains an average EER of 1.51\%, outperforming all baseline methods and indicating strong generalization capability.

The most significant improvement is observed on the ITW, which contains diverse real-world audio samples and is thus critical for evaluating cross-dataset generalization. As shown in Table~\ref{tab_inthewild}, our method achieves an EER of 4.70\%, corresponding to a 14.8\% relative improvement over the previous best result of 5.52\%. These results demonstrate the effectiveness of our method and its strong generalization capability in real-world scenarios.

Moreover, we use a radar chart to visualize the generalization ability of different models across four dimensions: performance on the 21LA, 21DF, and ITW, and the average performance across them. We adopt a ranking-based scoring strategy, assigning a score of 5 to the best method in each dimension and 1 to the worst. Unlike single-dataset results, this chart provides a more comprehensive comparison of generalization across multiple datasets. Four representative baseline methods \cite{Mamba, LSR+LSA, liu2025nes2net, 10446270} are selected for comparison. As shown in Fig.~\ref{fig-radar}, our method achieves the highest overall scores on the radar chart. This performance underscores our model’s generalization ability across different datasets.

ASVspoof5 represents the most recent edition in the ASVspoof series. Under this evaluation protocol, the model is trained on the official ASVspoof5 training set and evaluated on the corresponding test set. Table~\ref{tab:asv5} reports the performance comparison on this dataset. Our method achieves the best results among the compared systems in terms of CLLR, minDCF, and EER, outperforming the Nes2Net-X with relative improvements of approximately 16.6\% in EER. These results further demonstrate the effectiveness of our proposed approach on newly introduced spoofing conditions.

\begin{table}[t]
\centering
\caption{EER (\%) performance for ablation modules on eval datasets.}
\label{tab:ablation}
\setlength{\tabcolsep}{2.5mm}
{
\begin{tabular}{lccccc}
\hline
\toprule
Module                                                            & 19LA  & 21LA  & 21DF  & ITW   & Avg.\ \\
\hline
\midrule
w/o BreathFiLM                                                    & 0.30  & 1.06  & 2.19  & 6.16  & 3.43 \\
w/o Freq. Feat                                                    & \textbf{0.20}  & 1.34  & 2.05  & 6.01  & 2.40 \\
w/o $\mathcal{L}_{\text{feature}}$                                 & 0.30  & \textbf{0.98}  & 2.92  & 6.20  & 2.60 \\
w/o $\mathcal{L}_{\text{PSCL}}$                                    & 0.28  & 1.66  & 2.48  & 5.21  & 2.41 \\
w/o $\mathcal{L}_{\text{center}}$                                 & 0.27  & \textbf{0.98}  & 2.47  & 5.68  & 2.35 \\
w/o $\mathcal{L}_{\text{contrast}}$                               & 0.31  & 1.02  & 3.56  & 8.13  & 3.26 \\
\midrule
\rowcolor{gray!30} Default  & 0.23  & 1.15  & \textbf{1.87}  & \textbf{4.70}  & \textbf{1.99} \\
\bottomrule
\hline
\end{tabular}
}
\end{table}

\begin{figure*}
	\captionsetup[subfigure]{labelformat=empty}
	\centering
	\subfloat[(a) $\lambda$\label{fig-lambda}]{\includegraphics[width=0.33\textwidth]{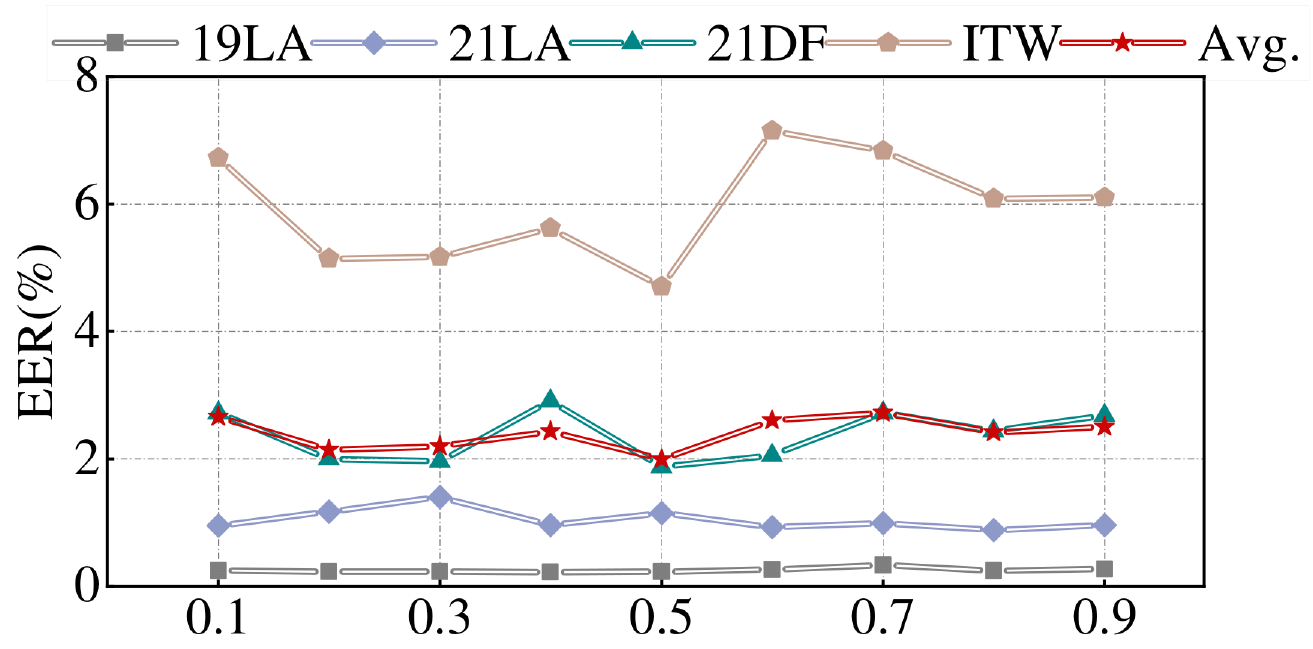}}
	\subfloat[(b) $\alpha$\label{fig-alpha}]{\includegraphics[width=0.33\textwidth]{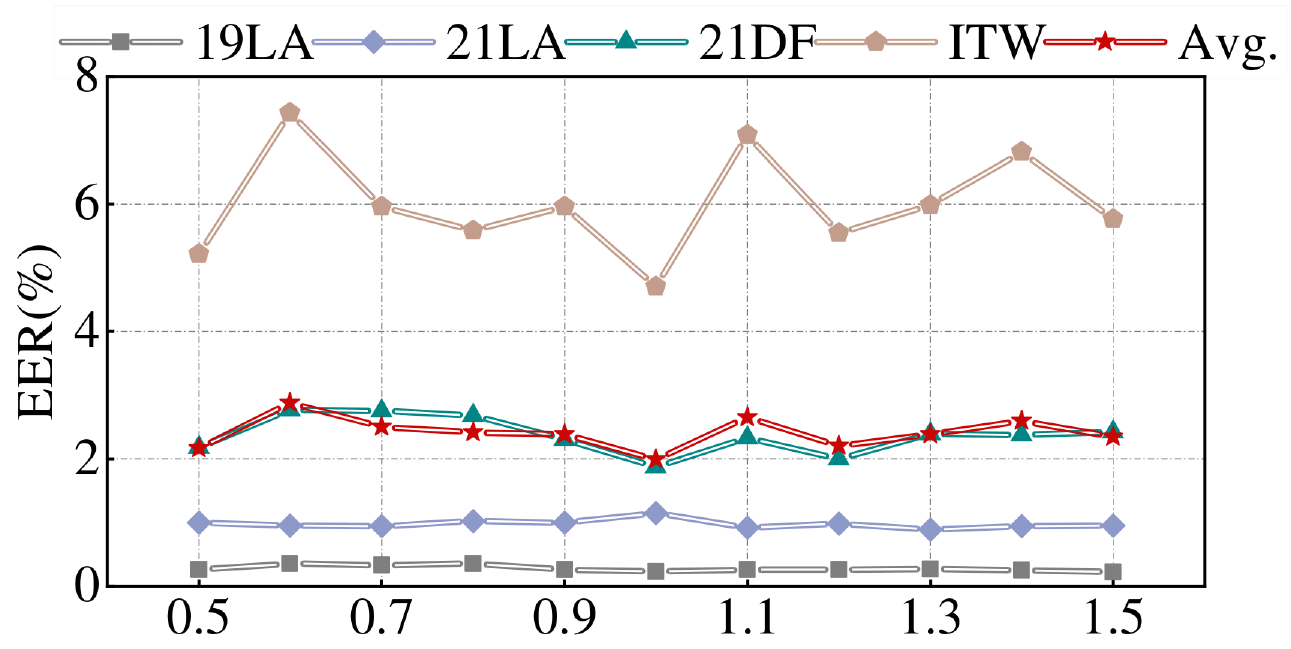}}
	\subfloat[(c) $\beta$\label{fig-beta}]{\includegraphics[width=0.33\textwidth]{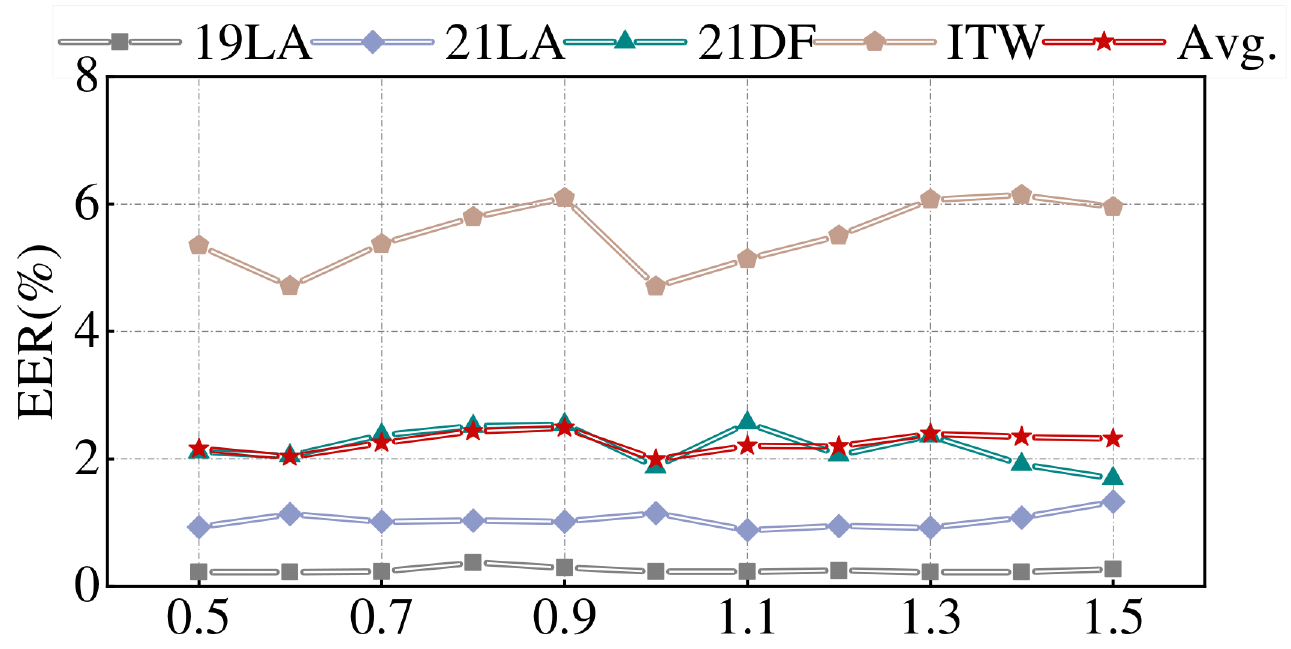}}
	\caption{EER (\%) performance for three feature loss on eval datasets.}
	\label{fig-loss}
\end{figure*}

\begin{figure*}
	\captionsetup[subfigure]{labelformat=empty}
	\centering
	\subfloat[(a) Number\label{fig-Number}]{\includegraphics[width=0.33\textwidth]{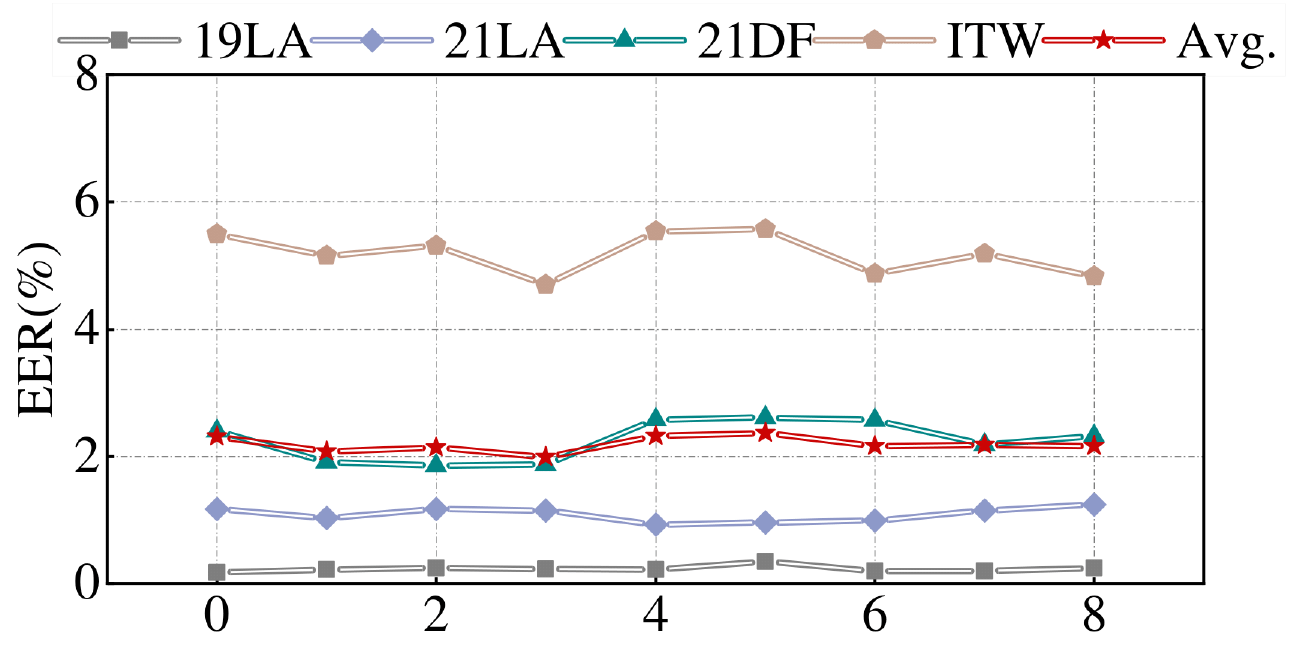}}
	\subfloat[(b) $\delta$\label{fig-delta}]{\includegraphics[width=0.33\textwidth]{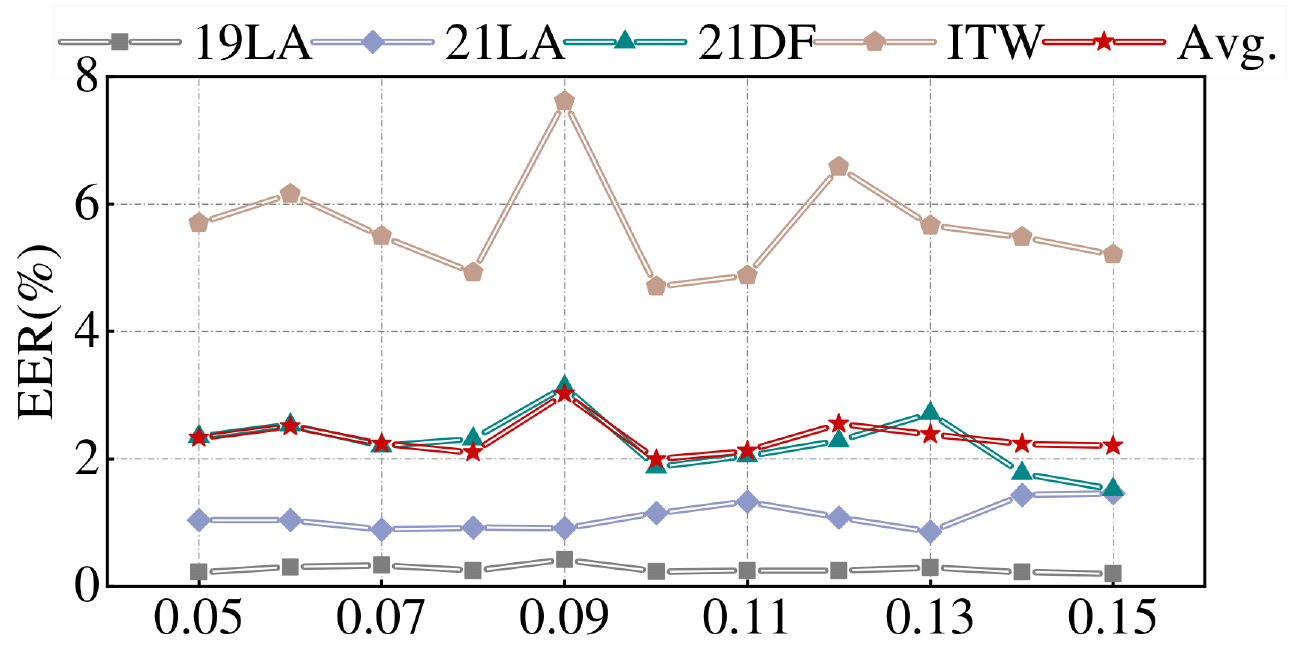}}
	\caption{EER (\%) performance for augmenting noise number and intensity on eval datasets.}
	\label{fig-sclnoise}
\end{figure*}

\subsection{Ablation Study}

\subsubsection{Ablation Modules}

We first conduct a qualitative ablation study to analyze the contribution of each component in our proposed method. Table~\ref{tab:ablation} reports the EER of our model on 19LA, 21LA, 21DF, and ITW datasets when key modules are removed. It is evident that removing any single module results in a degradation of average performance. The contribution of the BreathFiLM module is particularly significant, as its removal leads to an increase in EER from 0.23\% to 0.30\% on 19LA, from 1.87\% to 2.19\% on 21DF, and from 4.7\% to 6.16\% on ITW. The only exception is 21LA, where the performance remains similar. These results confirm the importance of capturing breath-related cues for detecting deepfake audio. Moreover, we observe that frequency-domain features provide valuable complementary information to temporal features. Without the frequency feature, the EER on 21LA, 21DF, and ITW increases by 0.19\%, 0.18\%, and 1.31\%, respectively. 

Besides, we validate the effectiveness of our feature loss. Removing the $\mathcal{L}_{\text{feature}}$ causes a noticeable performance drop, with the average EER increasing from 1.99\% to 2.60\%. A more detailed analysis of its three loss components further confirms their individual contributions: excluding the $\mathcal{L}_{\text{PSCL}}$ increases the average EER to 2.41\%, while removing the $\mathcal{L}_{\text{center}}$ or the $\mathcal{L}_{\text{contrast}}$ increases it to 2.35\% and 3.26\%, respectively. These results suggest that all three loss components make meaningful contributions. When combined, they yield the best discriminative capability by simultaneously enhancing intra-class compactness and inter-class separability.

Then, we conduct a quantitative ablation study on the feature loss to evaluate the contribution of each component individually.

\textbf{Feature Loss.} 
To evaluate the performance of our feature loss on different datasets, we conduct experiments with varying weight coefficients $\lambda$ from 0.1 to 1. As shown in Fig.~\ref{fig-lambda}, $\lambda = 1$ achieves the best performance on 19LA with an EER of 0.19\%, and achieves 1.72\% on 21DF. When $\lambda = 0.8$, the model reaches its best performance on 21LA with an EER of 0.88\%. Notably, setting $\lambda = 0.5$ results in the best performance on the ITW with an EER of 4.70\%, while also maintaining competitive performance across the other three datasets, leading to the best average performance.

\textbf{Center Loss.}
To evaluate the effect of the center loss coefficient $\alpha$ on model performance, we conduct experiments by varying $\alpha$ from 0.5 to 1.5. As shown in Fig.~\ref{fig-alpha}, setting $\alpha = 1$ gets the best EER on 21DF and ITW. When $\alpha = 1.5$, we observed the best performance on 19LA, resulting in an EER of 0.22\%. For 21LA, the best results are obtained with $\alpha = 1.3$. 

In addition, we observe that under the majority of center loss coefficients $\alpha$, the EER on 21LA is below 1\%, whereas the EER on 21DF typically exceeds 2\%. This discrepancy can be attributed to the inherent differences between the two datasets: 21LA focuses more on channel and codec variations, while 21DF emphasizes the diversity of spoofing algorithms. These findings suggest that the center loss is particularly effective at modeling and reducing distortions caused by transmission channels, making it more suitable for scenarios where channel variability is a dominant factor.

\textbf{Contrast Loss.}
We investigate the impact of the contrast loss coefficient $\beta$ by varying its value from 0.5 to 1.5. As shown in Fig.~\ref{fig-beta}, setting $\beta = 1$ provides the best average performance. Notably, setting $\beta = 1.5$ yields the lowest EER 1.69\% on 21DF. For 21LA, the best performance of 0.88\% is obtained when $\beta = 1.1$, while multiple values yield the lowest EER of 0.22\% on 19LA.

\begin{table*}[t]
\centering
\caption{EER (\%) performance for different model structures on eval datasets.}
\label{tab:model}
\setlength{\tabcolsep}{5.2mm}
{
\begin{tabular}{cccccccc}
\hline
\toprule
Model            & Hidden   & dropout & 19LA & 21LA  & 21DF  & ITW   & Avg.\ \\
\hline
\midrule
% Back-end, layer=2, Hidden=512-256, 4 种 dropout
\multirow{6}{*}{Back-end (layer=1)}    & 512-512 & -  & 0.23 & 1.04  & 2.57  & 5.25  & 2.27 \\
\rowcolor{gray!30} \cellcolor{white} 
                                       & 512-256 & -  & 0.23 & 1.15  & \textbf{1.87}  & \textbf{4.70}  & \textbf{1.99} \\ 
                                       & 512-128 & -  & 0.27 & 1.14  & 2.29  & 5.01  & 2.18 \\
                                       & 1024-256& -  & \textbf{0.19} & 1.18  & 1.99  & 5.32  & 2.17 \\
                                       & 256-256 & -  & 0.24 & 1.43  & 2.08  & 11.10 & 3.71 \\
                                       & 128-256 & -  & 0.45 & 1.11  & 3.56  & 8.95  & 3.44 \\
\midrule

\multirow{4}{*}{Back-end (layer=2)}  
                 & \multirow{8}{*}{512-256} & -    & 0.22 & 0.97  & 1.88  & 5.25  & 2.08 \\
                 &                           & 0.1  & 0.27 & \textbf{0.88}  & 2.13  & 5.61  & 2.23 \\
                 &                           & 0.2  & 0.23 & 0.97  & 2.91  & 5.45  & 2.39 \\
                 &                           & 0.3  & 0.26 & 1.15  & 2.27  & 5.57  & 2.31 \\

\multirow{4}{*}{Back-end (layer=3)}  
                 &                           & -    & 0.41 & 1.11  & 2.39  & 7.69  & 2.90 \\
                 &                           & 0.1  & 0.20 & 1.16  & 1.93  & 6.36  & 2.41 \\
                 &                           & 0.2  & 0.24 & 1.32  & 1.98  & 5.60  & 2.29 \\
                 &                           & 0.3  & 0.34 & 1.13  & 2.27  & 7.79  & 2.88 \\

\hline
\midrule
\multirow{5}{*}{BreathFiLM} 
                                       & 32  & -  & 0.41 & 1.03  & 2.54  & 5.51  & 2.37 \\
                                       & 64  & -  & 0.27 & 1.04  & 2.10  & 5.46  & 2.22 \\
                                       & 128 & -  & 0.27 & \textbf{0.90}  & 2.33  & 5.68  & 2.30 \\
                                       & 256 & -  & \textbf{0.19} & 1.19  & 1.96  & 4.76  & 2.03 \\
\rowcolor{gray!30} \cellcolor{white}   & 512 & -  & 0.23 & 1.15  & \textbf{1.87}  & \textbf{4.70}  & \textbf{1.99} \\ 
\bottomrule
\hline
\end{tabular}
}
\end{table*}

\begin{table}[t]
\centering
\caption{EER (\%) performance for different audio durations on eval datasets.}
\label{tab:duration}
\setlength{\tabcolsep}{2.8mm}
{
\begin{tabular}{cccccc}
\hline
\toprule
Durations (s)  & 19LA  & 21LA  & 21DF  & ITW   & Avg.\                           \\
\hline
\midrule
1          & 10.12  & 13.07  & 13.47  & 26.95  &  15.90                      \\
2          & 1.40  & 2.83  & 2.37  & 6.29  &  3.23                           \\
3          & 1.00  & 1.36  & 2.27  & 5.01  &  2.41                           \\
5          & \textbf{0.20}  & \textbf{0.77}  & 2.99  & 7.18  &  2.79         \\
6          & \textbf{0.20}  & 1.15  & 2.00  & 4.72  &  2.02                  \\
\midrule
\rowcolor{gray!30} 4  & 0.23  & 1.15  & \textbf{1.87}  & \textbf{4.70}  & \textbf{1.99} \\
\bottomrule
\hline
\end{tabular}
}
\end{table}

\textbf{PSCL.} 
We evaluate the effect of augmentation in the PSCL with respect to the number of augmented bona fide features per batch and the noise intensity $\delta$ applied to them.

First, we vary the number of augmented features from 0 to 9. As shown in Fig.~\ref{fig-Number}, using 3 augmented features yields the best average performance across all datasets. The best results on 19LA, 21LA, and 21DF are achieved when the number of augmented features is set to 0, 4, and 2, respectively. Second, we investigate the impact of the noise intensity $\delta$ by varying its standard deviation from 0.05 to 0.15. As shown in Fig.~\ref{fig-delta}, setting the noise standard deviation to 0.10 leads to the best average performance. It is also worth noting that if we exclude the ITW dataset and focus only on the ASVspoof series, the best performance is observed when the noise standard deviation is set to 0.15. Under this setting, the model achieves EER of 0.20\% on 19LA, 1.46\% on 21LA, and 1.52\% on 21DF.

\subsubsection{Back-end and BreathFiLM Model}

To explore the impact of model structure on detection performance, we evaluate different configurations for the back-end classifier (BiLSTM) and the BreathFiLM module. 

As shown in Table~\ref{tab:model}, the configuration of the back-end classifier (BiLSTM) and the BreathFiLM module has a significant impact on detection performance. For the BiLSTM, we vary the number of layers, hidden dimensions, and dropout rates. The best performance is achieved with a single-layer BiLSTM using hidden dimensions of 512 and 256.

For the BreathFiLM module, we change the hidden dimensions of its internal MLP. The results show that using a hidden size of 512 yields the best average results, while smaller hidden configurations lead to a marked drop in performance. This suggests that a larger modulation capacity is necessary for capturing subtle breath-related cues in the temporal features.

\begin{figure*}
	\captionsetup[subfigure]{}
	\centering
	\subfloat[19LA (spoof)]{\includegraphics[width=0.25\textwidth]{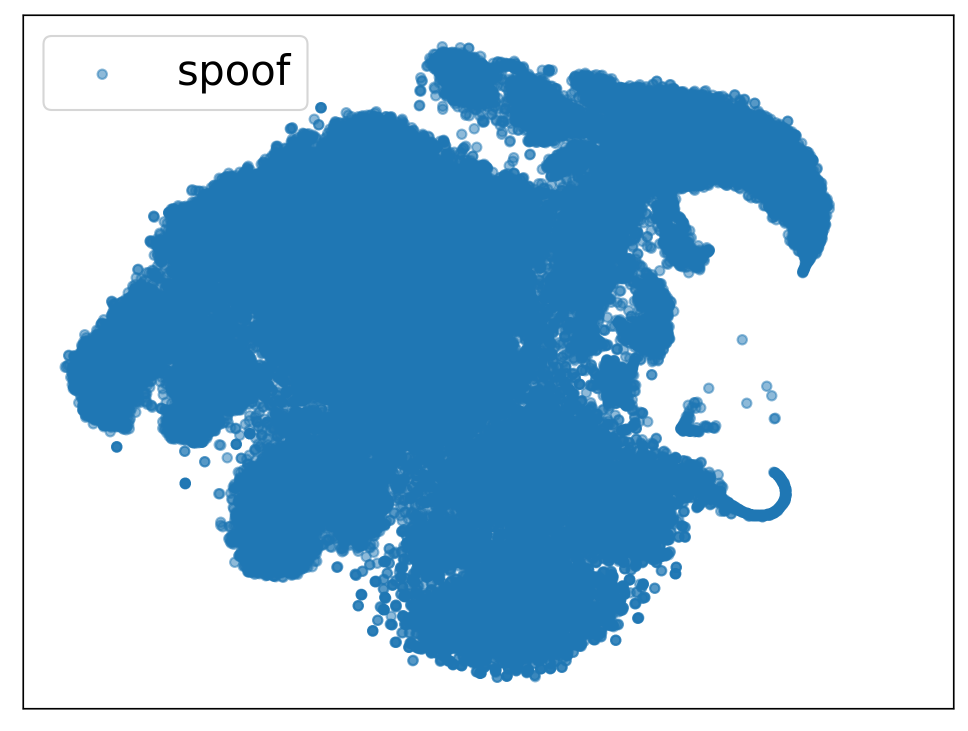}}
	\subfloat[21LA (spoof)]{\includegraphics[width=0.25\textwidth]{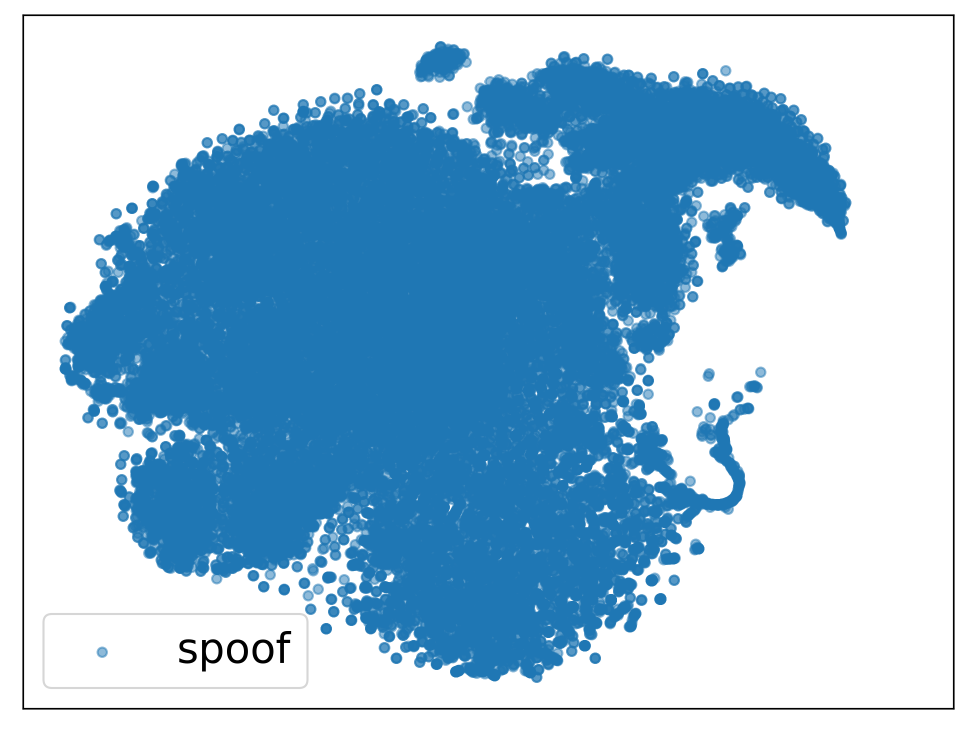}}
	\subfloat[21DF (spoof)]{\includegraphics[width=0.25\textwidth]{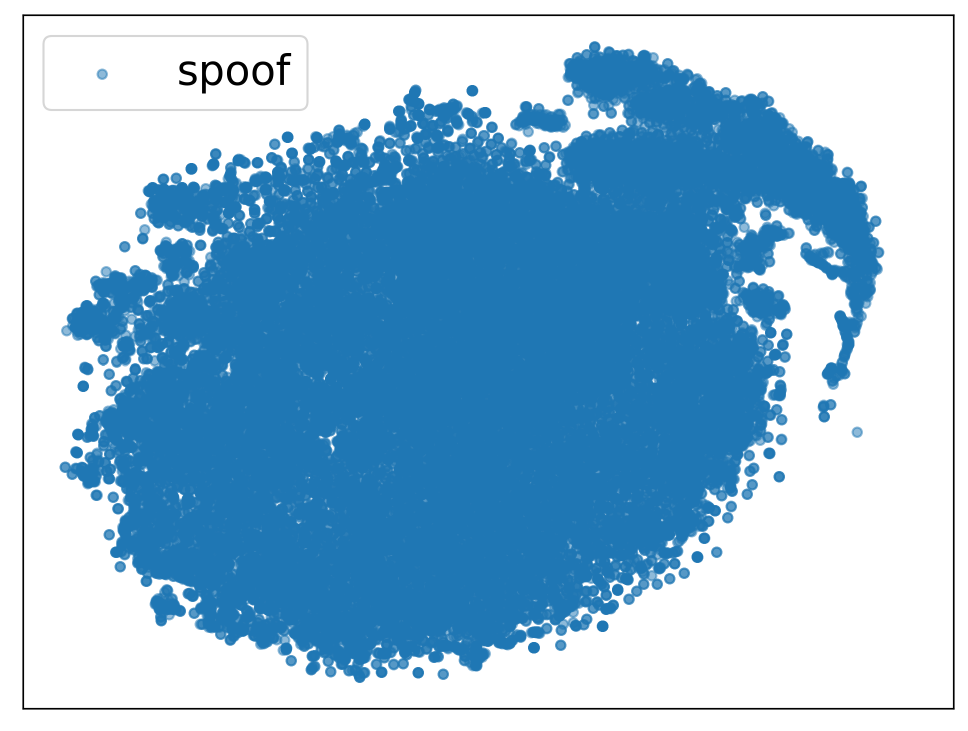}}
	\subfloat[ITW (spoof)]{\includegraphics[width=0.25\textwidth]{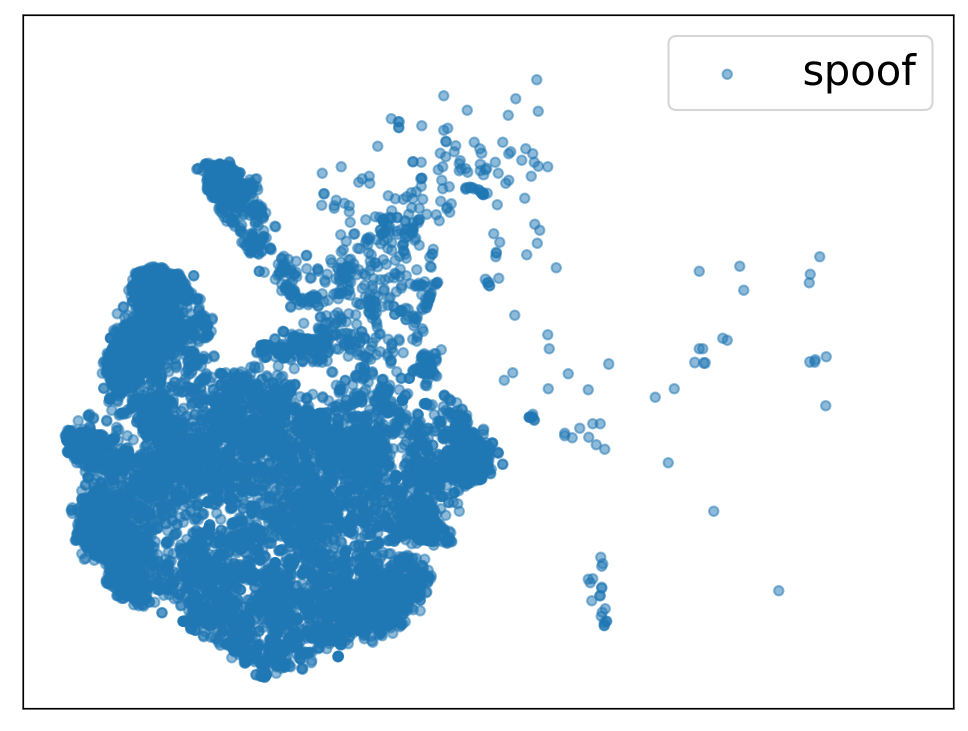}}
        \\
	\subfloat[19LA (bonafide)]{\includegraphics[width=0.25\textwidth]{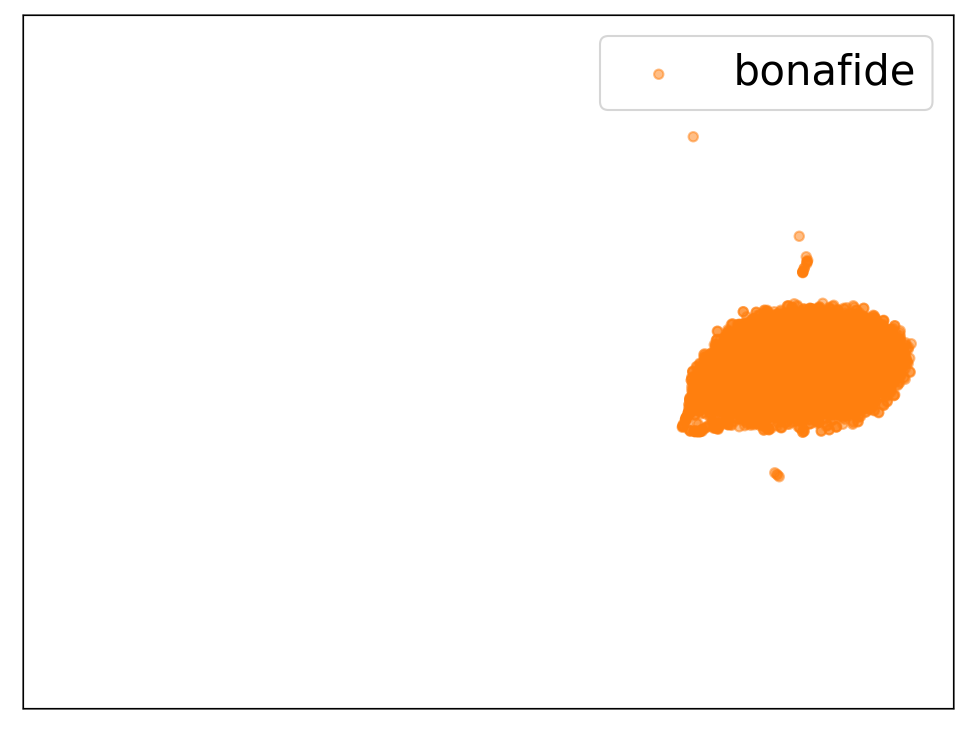}}
	\subfloat[21LA (bonafide)]{\includegraphics[width=0.25\textwidth]{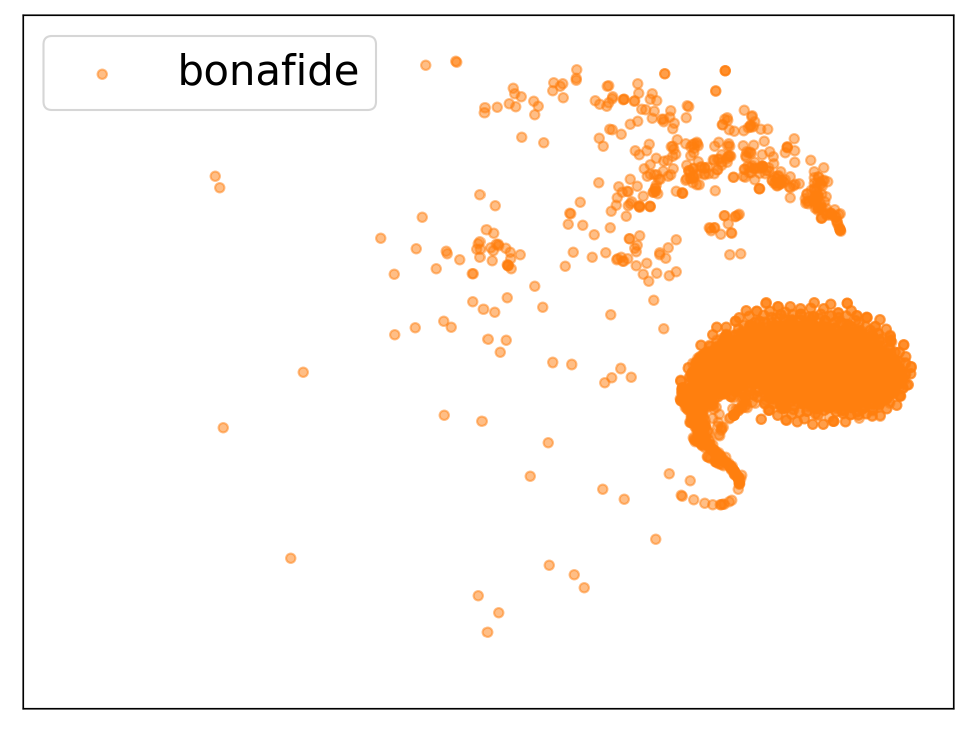}}
	\subfloat[21DF (bonafide)]{\includegraphics[width=0.25\textwidth]{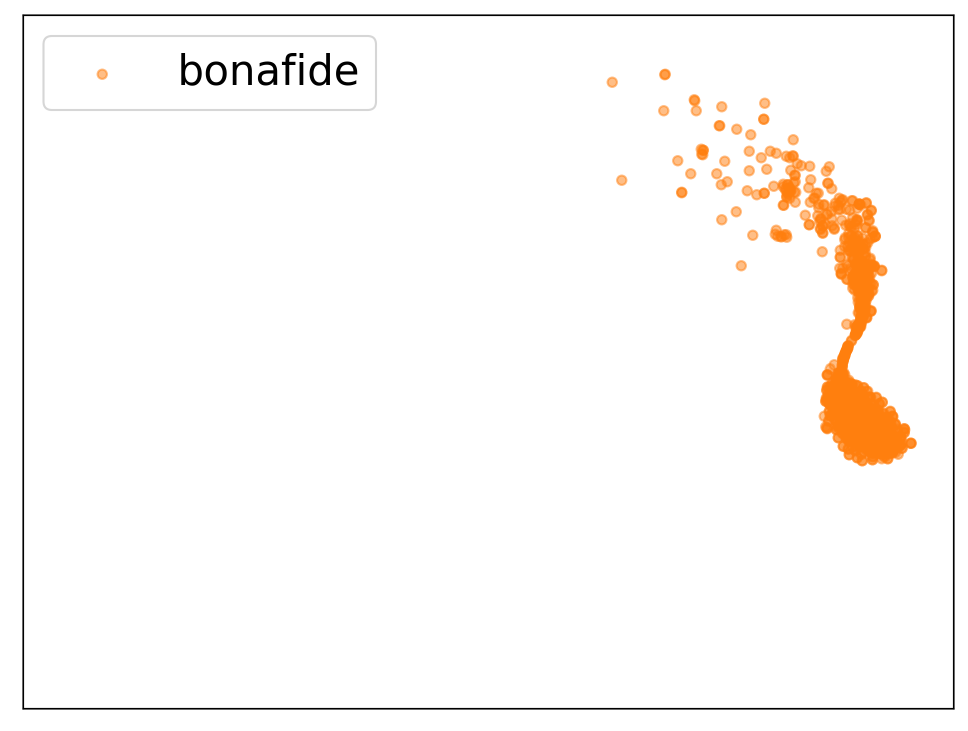}}
	\subfloat[ITW (bonafide)]{\includegraphics[width=0.25\textwidth]{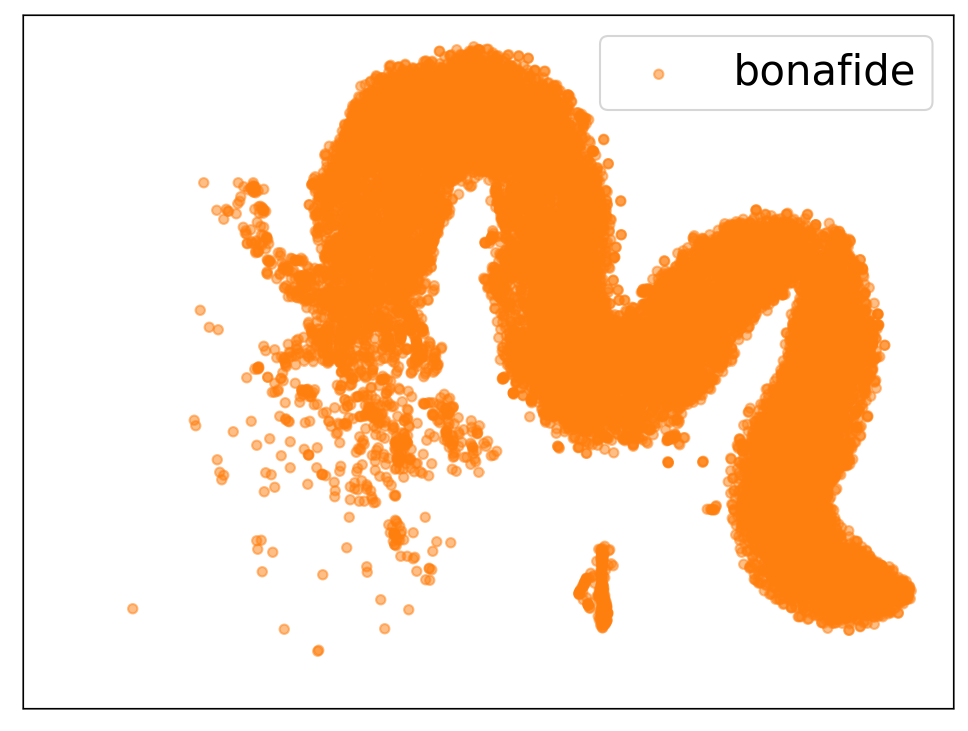}}
	\caption{t-SNE visualization of the hidden features in eval set.}
	\label{fig-tsne}
\end{figure*}

\begin{table*}[htbp!]
\centering
\setlength{\tabcolsep}{1.5mm}
\caption{EER (\%) Results of Ours/SLS/XLSR-MAMBA on the 21LA. Each column corresponds to a specific channel codec condition, including:  LA-C1: (nocodec, 16 kHz, 250 kbps), LA-C2: (a-law, 8 kHz, VoIP, 64 kbps), LA-C3: (unk. + $\mu$-law, 8 kHz, PSTN + VoIP, - / 64 kbps), LA-C4: (G.722, 16 kHz, VoIP, 64 kbps), LA-C5: ($\mu$-law, 8 kHz, VoIP, 64 kbps), LA-C6: (GSM, 8 kHz, VoIP, 13 kbps), LA-C7: (OPUS, 16 kHz, VoIP, VBR 16 kbps). Each row (A07-A19) corresponds to a different deepfake method.}
\label{tab:21la-}
\begin{tabular}{lcccccccc}

\toprule
& LA-C1  & LA-C2 & LA-C3 & LA-C4 & LA-C5 & LA-C6 & LA-C7 & Pooled \\
\hline
\toprule
A07 & \textbf{0.00}/\textbf{0.00}/0.06 & 0.62/\textbf{0.22}/0.37 & 0.24/\textbf{0.15}/0.21 & 0.67/\textbf{0.15}/0.52 & 0.48/\textbf{0.13}/0.38 & 1.16/\textbf{0.18}/0.85 & 1.35/\textbf{0.18}/0.80 & 0.77/\textbf{0.18}/0.46 \\
A08& \textbf{0.00}/\textbf{0.00}/0.21 & 0.57/\textbf{0.36}/0.45 & 0.28/0.43/\textbf{0.06} & 0.58/\textbf{0.30}/0.60 & 0.43/\textbf{0.20}/0.43 & 0.98/\textbf{0.55}/0.80 & 1.15/\textbf{0.37}/0.98 & 0.67/\textbf{0.41}/0.61\\

A09 & \textbf{0.00}/\textbf{0.00}/\textbf{0.00} & 0.37/\textbf{0.06}/0.30 & 0.15/\textbf{0.06}/\textbf{0.06} & 0.55/\textbf{0.00}/0.46 & 0.30/\textbf{0.00}/0.32 & 0.50/\textbf{0.07}/0.63 & 0.38/\textbf{0.00}/0.61 & 0.38/\textbf{0.05}/0.35\\

A10 & 0.15/0.88/\textbf{0.09} & 0.80/1.98/\textbf{0.46} & 0.67/3.19/\textbf{0.37} & 0.73/1.61/\textbf{0.61} & 0.62/1.83/\textbf{0.48} & 2.25/2.33/\textbf{1.05} & 1.78/3.10/\textbf{1.10} & 1.28/5.93/\textbf{0.72}\\

A11 & 0.09/0.91/\textbf{0.06} & 0.74/1.41/\textbf{0.46} & 0.58/2.27/\textbf{0.24} & 0.67/1.35/\textbf{0.61} & 0.60/1.46/\textbf{0.48} & 1.78/1.48/\textbf{0.90} & 1.60/3.48/\textbf{1.05} & 1.06/4.79/\textbf{0.63}\\

A12 & \textbf{0.00}/0.30/0.06 & 0.74/0.83/\textbf{0.43} & 0.36/0.58/\textbf{0.06} & 0.67/0.76/\textbf{0.58} & 0.61/0.80/\textbf{0.38} & 1.53/0.55/\textbf{0.86} & 1.61/1.36/\textbf{0.88} & 1.01/1.76/\textbf{0.52}\\

A13 & \textbf{0.00}/\textbf{0.00}/\textbf{0.00} & 0.37/\textbf{0.00}/0.30 & 0.06/\textbf{0.00}/0.06 & 0.46/\textbf{0.00}/0.46 & 0.35/\textbf{0.00}/0.37 & 0.60/\textbf{0.00}/0.78 & 0.43/\textbf{0.00}/0.60 & 0.37/\textbf{0.00}/0.40\\

A14 & \textbf{0.00}/\textbf{0.00}/0.06 & 0.52/\textbf{0.22}/0.37 & 0.21/\textbf{0.09}/0.15 & 0.59/\textbf{0.06}/0.52 & 0.43/\textbf{0.13}/0.38 & 1.03/\textbf{0.18}/0.91 & 0.98/\textbf{0.35}/0.80 & 0.65/\textbf{0.19}/0.47\\

A15 & \textbf{0.00}/0.06/0.06 &  0.61/0.61/\textbf{0.31} & 0.30/0.36/\textbf{0.09} & 0.67/0.67/\textbf{0.52} & 0.48/0.50/\textbf{0.38} & 1.31/\textbf{0.50}/0.86 & 1.43/\textbf{0.48}/0.81  & 0.84/0.73/\textbf{0.47}\\

A16 & \textbf{0.06}/\textbf{0.06}/\textbf{0.06} & 0.64/0.64/\textbf{0.45} & 0.45/0.67/\textbf{0.30} & 0.67/0.67/\textbf{0.60} & 0.50/0.50/\textbf{0.48} & 1.45/1.10/\textbf{0.85} & 1.45/0.97/\textbf{0.90} & 0.86/1.00/\textbf{0.59}\\

A17 & 0.45/0.97/\textbf{0.15} & 1.19/1.55/\textbf{0.67} & 1.04/4.35/\textbf{0.37} & 0.89/1.13/\textbf{0.61} & 0.98/1.41/\textbf{0.48} & 1.96/7.34/\textbf{1.05} & 1.36/2.34/\textbf{0.98} & 1.21/3.47/\textbf{0.71}\\

A18 & 1.55/\textbf{1.27}/1.79 & 2.34/1.72/\textbf{1.35} & 2.52/5.32/\textbf{1.45} & 1.87/\textbf{1.44}/1.87 & 1.87/1.74/\textbf{1.36} & 2.85/5.70/\textbf{1.69} & 2.42/2.42/\textbf{1.99} & 2.75/3.63/\textbf{2.60}\\

A19 & \textbf{0.52}/0.88/0.61 & \textbf{0.98}/1.50/1.13 & 1.20/4.74/\textbf{1.08} & \textbf{0.89}/1.19/1.13 & \textbf{0.86}/1.28/0.93 & 1.96/8.27/\textbf{1.43} & 1.84/2.41/\textbf{1.73} & \textbf{1.43}/3.69/1.68\\

Pooled & \textbf{0.32}/0.51/0.46 & 0.78/1.15/\textbf{0.73} & 0.62/2.38/\textbf{0.47} & \textbf{0.72}/1.08/0.83 & \textbf{0.63}/1.10/0.65 & 1.87/3.48/\textbf{0.93} & 1.61/2.16/\textbf{1.19} & 1.15/2.87/\textbf{0.93}\\
\hline
\end{tabular}
\end{table*}

\begin{table*}[htbp!]
\centering
\setlength{\tabcolsep}{3mm}
\caption{EER (\%) Results of Ours/SLS/XLSR-MAMBA on the 21DF. Each row corresponds to specific compression condition, including: DF-C1 (no compression, 256 kbps), DF-C2 (Low mp3, ~80–120 kbps), DF-C3 (High mp3, ~220–260 kbps), DF-C4 (Low m4a, ~20–32 kbps), DF-C5 (High m4a, ~96–112 kbps), DF-C6 (Low ogg, ~80–96 kbps), DF-C7 (High ogg, ~256–320 kbps), DF-C8 (mp3$\rightarrow$m4a, ~80–120 kbps and ~96–112 kbps), and DF-C9 (ogg$\rightarrow$m4a, ~80–96 kbps and ~96–112 kbps). Each column corresponds to a specific generative vocoder.}
\label{tab:21df-}
\begin{tabular}{lcccccc}

\toprule
& Traditional Vocoder & Wav Concatenation & Neural AR & Neural non-AR & Unknown & Pooled \\
\hline
\toprule
DF-C1  & \textbf{0.61}/1.21/0.78 & \textbf{0.76}/0.80/0.76 & 4.32/\textbf{3.12}/3.88 & \textbf{0.61}/0.68/0.87 & 2.06/\textbf{1.23}/1.63 & 1.99/\textbf{1.72}/1.89 \\
DF-C2  & \textbf{0.70}/1.94/0.94 & \textbf{1.48}/2.16/2.20 & 2.98/\textbf{2.71}/3.23 & \textbf{0.62}/0.78/0.86 & \textbf{1.56}/1.65/1.69 & \textbf{1.49}/2.02/1.84 \\
DF-C3  & \textbf{0.54}/1.39/0.88 & \textbf{1.04}/1.17/1.49 & 3.36/\textbf{2.91}/3.35 & \textbf{0.51}/0.69/0.87 & 2.22/\textbf{1.34}/1.85 & 1.85/\textbf{1.59}/1.85 \\
DF-C4  & \textbf{0.61}/1.48/0.95 & \textbf{0.43}/1.24/0.85 & 3.49/\textbf{2.79}/3.39 & \textbf{0.59}/0.70/0.96 & 1.40/\textbf{1.14}/1.22 & \textbf{1.74}/\textbf{1.74}/1.92 \\
DF-C5  & \textbf{0.63}/1.34/0.80 & \textbf{0.71}/\textbf{0.71}/0.76 & 3.86/\textbf{2.96}/3.48 & 0.79/\textbf{0.64}/0.90 & 1.96/\textbf{1.34}/1.70 & 1.88/\textbf{1.79}/2.05 \\
DF-C6  & \textbf{1.00}/2.14/1.13 & \textbf{0.62}/0.91/0.97 & 2.67/\textbf{2.44}/2.80 & \textbf{0.57}/0.61/0.78 & \textbf{0.96}/1.00/1.14 & \textbf{1.61}/1.88/\textbf{1.61} \\
DF-C7  & \textbf{1.00}/1.52/1.13 & \textbf{0.46}/0.71/0.80 & 3.59/\textbf{2.26}/2.84 & \textbf{0.48}/0.52/0.65 & 1.58/\textbf{0.96}/1.05 & 1.88/\textbf{1.57}/1.61 \\
DF-C8 & \textbf{1.09}/2.28/1.26 & \textbf{0.80}/1.08/0.97 & 2.84/\textbf{2.31}/3.01 & \textbf{0.44}/0.65/0.57 & 1.18/\textbf{1.09}/1.18 & \textbf{1.62}/1.92/1.65 \\
DF-C9 & \textbf{1.00}/2.15/1.26 & \textbf{0.55}/0.99/0.97 & 2.92/\textbf{2.57}/3.01 & \textbf{0.52}/0.65/0.70 & \textbf{0.96}/1.09/1.09 & \textbf{1.87}/2.04/1.79 \\

Pooled & \textbf{0.91}/1.88/1.14 & \textbf{0.77}/1.07/1.05 & 3.54/\textbf{2.86}/3.32 & \textbf{0.62}/0.69/0.80 & 1.67/\textbf{1.23}/1.43 & \textbf{1.87}/1.92/1.88 \\
\hline
\end{tabular}
\end{table*}

\subsubsection{Sample Durations}
To investigate the impact of audio duration on the detection performance of our model, we evaluate our model using audio durations from 1 to 6 seconds. Following standard practice, each input is constructed by taking the first $n$ seconds of the sample. If the sample is shorter than $n$, it is padded by repeating the original audio until the desired length is reached. As shown in Table~\ref{tab:duration}, 1-second audios fail to provide sufficient contextual information for detection. Although 2-second audios perform poorly on 19LA and 21LA, they already achieve acceptable performance on 21DF and ITW. The performance on 21LA becomes acceptable with 3-second segments. Overall, 4 and 6-second audios lead to better performance across four datasets, suggesting that longer audio durations better support our model’s ability for deepfake detection.

% \subsubsection{XLS-R Layers}
% To determine the most effective weighted aggregation strategy for XLS-R, we evaluate the performance using different layer outputs. As shown in Table~\ref{tab:layers}, applying a weighted aggregation of all 24 transformer layers achieves the best results. These results indicate that each transformer layer in XLS-R contributes unique discriminative information, and leveraging all layers enables more comprehensive feature representation for audio deepfake detection.

\subsection{t-SNE Analysis}

In this section, we analyze the latent feature representations learned by the model using t-SNE. Specifically, we extracted the final hidden features from the evaluation sets. Due to the large size of the 21LA and 21DF, we randomly selected 40,000 samples for visualization, while the others used all samples. 

As shown in Fig.~\ref{fig-tsne}, deepfake and bona fide samples on 19LA are generally well-separated into two distinct clusters. The sample distribution of 21LA closely resembles that of 19LA, but a portion of bona fide samples deviates from the main bona fide cluster and overlaps with the deepfake region. For 21DF, bona fide samples were split into two clusters, with one cluster significantly overlapping the deepfake region. This indicates that a portion of the bona fide samples in the 21DF is not well distinguished from deepfake samples, possibly due to the impact of codec compression or vocoder artifacts that blur the decision boundary. Notably, the ITW displays a reversed density pattern, with a larger and more continuous bona fide region compared to the deepfake region. Moreover, both deepfake and bona fide samples exhibit cross-cluster dispersion, indicating that the real-world distribution in ITW presents challenges for the model in clearly separating the classes. Overall, these visualizations highlight the discriminative capability of our model while also indicating potential directions for further optimization.

\subsection{Extra Analysis on 21LA and 21DF}
21LA and 21DF provide extra annotations for detailed evaluation: 21LA includes codec and attack algorithm labels, while 21DF offers vocoder types and compression conditions. In this section, we conduct an in-depth analysis of our method’s performance under various conditions in two datasets.

As shown in Table~\ref{tab:21la-}, our method achieves the best performance on 21LA under the codec conditions LA-C1, LA-C4, and LA-C5, while LA-C6 and LA-C7 are the most difficult conditions for our model. In terms of deepfake methods, our method performs well in detecting A19, while A18 is the most difficult to detect. For 21DF, as illustrated in Table~\ref{tab:21df-}, our method achieves the best performance under the Traditional Vocoder, Waveform Concatenation, and Neural Non-Autoregressive vocoder (Neural non-AR) conditions, with the Neural Autoregressive vocoder (Neural AR) posing the most significant challenge for our method. Furthermore, our method demonstrates superior robustness under multiple compression conditions, as seen in DF-C2, DF-C4, DF-C6, DF-C8, and DF-C9. 

% \begin{table}[t]
% \caption{The performance for Augmentation and Average Checkpoint on Evaluation Strategies.}
% \label{tab_evals}
% \centering
% \setlength{\tabcolsep}{0.3mm}
% {
% \begin{tabular}{cccccc|ccccc}
% \hline
% \toprule
% \multirow{2}{*}{\begin{tabular}[c]{@{}c@{}}CKPT \\ Avg.\end{tabular}} 
%     & \multicolumn{5}{c|}{w/o Aug.} 
%     & \multicolumn{5}{c}{w/ Aug.} \\
% \cmidrule(r){2-6} \cmidrule(r){7-11}
%     & 19LA & 21LA & 21DF & ITW & Avg.
%     & 19LA & 21LA & 21DF & ITW & Avg. \\
% \hline
% \midrule
% N/A & \textbf{0.17} & 1.78 & \textbf{1.62} & \textbf{4.33} & 1.98
%     & 0.22 & 2.65 & 2.48 & 6.46 & 2.45 \\
% 2   & 0.24 & 1.16 & 1.86 & 4.59 & \textbf{1.96}
%     & 0.26 & 1.53 & 2.83 & 6.83 & 2.86 \\
% \cellcolor{gray!30}3   & \cellcolor{gray!30}0.23 & \cellcolor{gray!30}\textbf{1.15} & \cellcolor{gray!30}1.87 & \cellcolor{gray!30}4.70 & \cellcolor{gray!30}1.99
%     & 0.25 & 1.47 & 2.75 & 6.97 & 2.86 \\
% \bottomrule
% \hline
% \end{tabular}
% }
% \end{table}

\begin{table}[t]
\caption{Comparison of four XLS-R training strategies. (1): Fine-tune pre-trained XLS-R, (2) Freeze pre-trained XLS-R, (3) Fine-tune self-trained XLS-R, (4) Freeze self-trained XLS-R. “pre-trained”  refers to the officially released XLS-R model. “self-trained” refers to an XLS-R model that has been previously trained using strategy (1).}
\label{tab:xlsr_strategies}
\centering
\setlength{\tabcolsep}{3mm}
{
\begin{tabular}{cccccc}
\hline
\toprule
Strategy & 19LA & 21LA & 21DF & ITW & Avg.\\
\hline
\midrule
\rowcolor{gray!30}(1)     & \textbf{0.23} & 1.15 & 1.87 & \textbf{4.70} & \textbf{1.99} \\
(2)     & 0.83          & 2.51          & 10.40         & 18.58         & 8.08          \\
(3)     & 0.26          & \textbf{0.94} & 1.88          & 6.25          & 2.33          \\
(4)     & 0.24          & 1.34          & \textbf{1.59} & 4.74          & \textbf{1.98} \\
\bottomrule
\hline
\end{tabular}
}
\end{table}

\subsection{Evaluation Strategies}

In this section, we investigate the impact of different evaluation strategies on model performance. As shown in Table~\ref{tab:xlsr_strategies}, Strategy (1) consistently outperforms Strategy (2), indicating that fine-tuning allows XLS-R to adapt its representations to the downstream deepfake detection task and provide more task-specific discriminative features. In contrast, freezing the pre-trained XLS-R results in significant performance degradation on challenging datasets, with EER of 10.40\% on 21DF and 18.58\% on ITW. Strategy (3) achieves the best performance on 21LA but performs poorly on 21DF and ITW, suggesting that further fine-tuning may overfit the model to the characteristics of 21LA, thereby reducing its generalizability to other datasets. Strategy (4) performs comparably to Strategy (1), demonstrating that once the XLS-R encoder is well fine-tuned for the deepfake detection task, it can be directly used without further fine-tuning.

\begin{table}[t]
\caption{Comparison of different breath mask strategies for BreathFiLM during inference. Normal denotes using the breath mask extracted from the test utterance itself, while the all zeros and all ones settings correspond to treating all frames as non-breath and breath regions, respectively.}
\label{tab:mask}
\centering
\setlength{\tabcolsep}{2.5mm}
{
\begin{tabular}{cccccc}
\hline
\toprule
Breath mask & 19LA & 21LA & 21DF & ITW & Avg.\\
\hline
\midrule
Normal    & 0.23 & 1.15 & 1.87 & 4.70 & 1.988 \\
All zeros & 0.23 & 1.16 & 1.86 & 4.71 & 1.990 \\
All ones  & 0.23 & 1.15 & 1.88 & 4.68 & 1.985 \\
\bottomrule
\hline
\end{tabular}
}
\end{table}

\section{Discussion} \label{Discussion}
From a practical standpoint, it is important to understand whether the proposed method depends on precise breath localization. Therefore, we analyze two extreme inference scenarios where the breath mask is uniformly assigned to all zeros or all ones, corresponding to treating all frames as non-breath or breath, respectively. By eliminating breath-related cues, this analysis allows us to examine whether the model’s decisions rely directly on the mask itself rather than on the learned features.

As shown in Table~\ref{tab:mask}, the EER remains nearly unchanged across 19LA, 21LA, 21DF, and ITW under these extreme mask conditions compared to the normal mask setting. This result indicates that our model is not making decisions based on the presence of breathing. Our BreathFiLM serves as an auxiliary mechanism that guides the fine-tuned XLS-R feature extractor to incorporate breath-related cues during training. Consequently, the feature extractor can implicitly encode such breath-related cues into the temporal features at inference, allowing the model to be evaluated without relying on the breath mask.

\section{Conclusion} \label{Conclusion}

In this paper, we propose BreathNet, a novel audio deepfake detection framework that enhances discriminability by incorporating frame-level breath cues into temporal–spectral features and designing a feature loss that improves intra-class compactness and inter-class separability. Importantly, while the breath mask is employed during training, it is not required at inference, thereby enhancing the practicality and deployability of our method. Extensive experiments on five datasets demonstrate the effectiveness and generalization ability of our method. Despite its promising performance, our analysis reveals that a portion of deepfake and bona fide samples still exhibit overlap in the feature space, particularly under complex conditions. To overcome it, more fine-grained and adaptive strategies need to be explored. In future work, we plan to extend our research in collaboration with China Mobile Internet Corporation to large-scale and diverse Chinese audio deepfake datasets, with a particular focus on real-world voice call scenarios, aiming to improve generalization and ensure reliable audio deepfake detection.

\clearpage
\bibliographystyle{IEEEtran}
\bibliography{main}

\end{document}